\documentclass[11pt,a4paper]{article}
\usepackage{amsmath,latexsym,fullpage,amssymb,color}
\usepackage{epic,eepic,ifthen,graphics,epsfig}
\usepackage[english]{babel}
\usepackage{times}
\usepackage{float}
\usepackage{xspace}
\usepackage[T1]{fontenc}
\usepackage{tikz}
\usetikzlibrary{decorations.markings}
\usetikzlibrary{decorations.pathmorphing}
\usetikzlibrary{decorations.pathreplacing}

\newlength {\squarewidth}

\newtheorem{definition}{Definition}
\newtheorem{theorem}{Theorem}
\newtheorem{lemma}{Lemma}
\newtheorem{corollary}{Corollary}

\newcommand{\toto}{xxx}
\newenvironment{proofT}{\noindent{\bf Proof }}
{\hspace*{\fill}$\Box_{Theorem~\ref{\toto}}$\par\vspace{3mm}}
\newenvironment{proofL}{\noindent{\bf Proof }}
{\hspace*{\fill}$\Box_{Lemma~\ref{\toto}}$\par\vspace{3mm}}

\newenvironment{lemma-repeat}[1]{\begin{trivlist}
\item[\hspace{\labelsep}{\bf\noindent Lemma~\ref{#1} }]}%
{\end{trivlist}}

\newcounter{linecounter}
\newcommand{\linenumbering}{\ifthenelse{\value{linecounter}<10}
{(\arabic{linecounter})}{(\arabic{linecounter})}}
\renewcommand{\line}[1]{\refstepcounter{linecounter}\label{#1}\linenumbering}
\newcommand{\resetline}[1]{\setcounter{linecounter}{0}#1}
\renewcommand{\thelinecounter}{\ifnum \value{linecounter} > 
9 \else \fi\arabic{linecounter}}
\newfloat{algorithm}{thp}{lop}
\floatname{algorithm}{Algorithm}
\newfloat{construction}{thp}{lop}
\floatname{construction}{Construction}
\newcommand{\Xomit}[1]{}

\newcommand{\REG}{\mathit{REG}}

\newcommand{\op}{{\sf op}}
\newcommand{\progress}{{\sf progress}}
\newcommand{\forward}{{\sf forward}}
\newcommand{\trydeliver}{{\sf try\_deliver}}
\newcommand{\ttrue}{\tt{true}}
\newcommand{\ffalse}{\tt{false}}
\newcommand{\done}{\mathit{done}}

\newcommand{\SYNC}{{\sc{sync}}}
\newcommand{\WRITE}{{\sc{write}}}
\newcommand{\FORWARD}{{\sc{forward}}}
\newcommand{\reg}{\mathit{reg}}
\newcommand{\buffer}{\mathit{buffer}}
\newcommand{\bufferi}{\mathit{buffer_i}}    
\newcommand{\bufferj}{\mathit{buffer_j}}

\newcommand{\snapshot}{{\sf snapshot}}
\newcommand{\wwrite}{{\sf write}}
\newcommand{\rread}{{\sf read}}

\newcommand{\catchup}{\sf{catch\_up}}
 
\newcommand{\wait}{{\sf{wait}}}
\newcommand{\return}{{\sf{return}}}

\newcommand{\mmin}{{\sf min}}

\newcommand{\snsd}{{\mathit{sn_{sd}}}}
\newcommand{\snf}{{\mathit{sn_{f}}}}

\newcommand{\SETSEQ}{\mathit{SETS\_SEQ}}
\newcommand{\setseq}{\mathit{sets\_seq}}
\newcommand{\SENT}{\mathit{SENT}}
\newcommand{\sent}{{\mathit{sent}}}

\newcommand{\members}{{\sf{members}}}

\newcommand{\entermutex}{\sf{enter\_mutex}}
\newcommand{\exitmutex}{\sf{exit\_mutex}}
\newcommand{\todeliver}{\mathit{to\_deliver}}
\newcommand{\todeliveri}{\mathit{to\_deliver_i}}

\newcommand{\CAMP}{{\cal{CAMP}}}
\newcommand{\CARW}{{\cal{CARW}}}
\newcommand{\SCD}{{{SCD}}}

\newcommand{\scbroadcast}{{\sf scd\_broadcast}}
\newcommand{\scdeliver}{{\sf scd\_deliver}}
\newcommand{\fifobroadcast}{{\sf fifo\_broadcast}}
\newcommand{\fifodeliver}{{\sf fifo\_deliver}}

\newcommand{\tsa}{\mathit{tsa}}
\newcommand{\TSA}{\mathit{TSA}}
\newcommand{\lexts}{{~<_{ts}~}}
\newcommand{\MS}{\mathit{MS}}

\newlength{\reduceunderfig}
\usepackage[lmargin=2.1cm,rmargin=2cm,tmargin=2cm,bmargin=2cm]{geometry}
\renewcommand{\baselinestretch}{0.96}
\renewcommand{\paragraph}[1]{\vspace{0.2cm}\noindent \textbf{#1}~~}
\let\subsectionOld\subsection
\renewcommand{\subsection}[1]{\vspace{-0.2cm}\subsectionOld{#1}\vspace{-0.2cm}}
\let\subsubsectionOld\subsubsection
\renewcommand{\subsubsection}[1]{\vspace{-0.3cm}\subsubsectionOld{#1}\vspace{-0.15cm}}

\begin{document}

\title{\bf  Another Look at the Implementation of  Read/write Registers \\
            in Crash-prone
            Asynchronous Message-Passing Systems \\
            (Extended Version)}

  \author{Damien Imbs$^{\circ}$, 
         Achour Most\'efaoui$^{\dag}$,~
         Matthieu Perrin$^{\vartriangle}$,~
         Michel Raynal$^{\star,\ddag}$\\~\\
$^{\circ}$LIF, Universit\'e Aix-Marseille, 13288  Marseille, France \\
$^{\dag}$LINA, Universit\'e de Nantes, 44322 Nantes, France \\
{$^{\vartriangle}$}Computer science department, Technion, Haifa, 3200003, Israel \\
$^{\star}$Institut Universitaire de France\\
$^{\ddag}$IRISA, Universit\'e de Rennes, 35042 Rennes, France \\
}

\date{}

\maketitle


\begin{abstract}
  ``Yet another paper on'' the implementation of read/write registers
  in crash-prone asynchronous message-passing systems! Yes..., but,
  differently from its predecessors, this paper looks for a
  communication abstraction which captures the essence of such an
  implementation in the same sense that total order broadcast can be
  associated with consensus, or message causal delivery can be
  associated with causal read/write registers.  To this end, the paper
  introduces  a new communication abstraction, named \SCD-broadcast
  (\SCD\ standing for ``Set Constrained Delivery''), which, instead of
  a single message, delivers to processes sets of messages (whose size
  can be arbitrary), such that the sequences of message sets delivered
  to any two processes satisfies some constraints.  The paper then
  shows that: (a) \SCD-broadcast allows for a very simple
  implementation of a snapshot object (and consequently also of atomic
  read/write registers) in crash-prone asynchronous message-passing
  systems; (b) \SCD-broadcast can be built from snapshot objects
  (hence \SCD-broadcast and snapshot objects --or read/write
  registers-- are ``computationally equivalent''); (c) \SCD-broadcast
  can be built in message-passing systems where any minority of
  processes may crash (which is the weakest assumption on the number
  of possible process crashes needed to implement a read/write
  register).
  
  ~
\end{abstract}


\thispagestyle{empty}
\newpage
\setcounter{page}{1}

\section{Introduction}

The ``one-shot'' terracotta tablets introduced and used at Sumer about 3030
BC~\cite{K56}, and the ``multi-shot'' palimpsests used in the middle-age,
can be considered as ancestors of the {\it read/write register}
abstraction. Such an object provides its users with a write operation
which defines a new value of the register, and a read operation which
returns its value.  When considering sequential computing, read/write
registers are universal in the sense that they are assumed to allow  solving
any problem that can be solved~\cite{T36}.

\paragraph{On the variety of read/write registers
           and their distributed implementation} 
In a shared read/write memory system, the registers are given for
free.  The situation is different in a message-passing system, where
the computing entities (processes) communicate by sending and
receiving messages transmitted through a communication network.
Hence, in such a distributed context, a register is not given for
free, but constitutes a communication abstraction which must be built by
a distributed algorithm with the help of the local memories of the
processes and the communication network.

Several types of registers have been proposed. They differ  according to
(a) their size (from binary registers  which contain a single bit,
to bounded and unbounded registers); (b) their behavior in the presence
of concurrency (safe, regular, atomic~\cite{L86}); 
(c) the number of processes
which are allowed to read  them (Single-Reader -SR- vs Multi-Reader -MR-
register);
and (d) the number of processes which are allowed to write them
(Single-Writer -SR- vs Multi-Writer -MR- register), which gives four
possible combinations from SWSR to MWMR. There are algorithms building
MWMR atomic (bounded and unbounded) registers from SWSR binary safe
registers~\cite{L86} (see~\cite{AW04,L96,R13-1} for surveys of such algorithms).

As far as a read/write register is concerned, {\it atomicity} means
that (a) each read or write operation appears as if it had been
executed instantaneously at a single point of the time line, (b) this
point appears between its start event and its end event, (c) no two
operations appear at the same point of the time line, and (d) a read
returns the value written by the closest preceding write operation (or
the initial value of the register if there is no preceding
write)~\cite{L86,M86}.  {\it Linearizability} is atomicity extended to
any object defined from a sequential specification on total
operations~\cite{HW90}.  In the following, we consider the terms
atomicity and linearizability as synonyms.  Hence, a sequence of read
and write operations satisfying atomicity is said to be linearizable,
and is called a linearization. The point of the time line at which an
operation appears to have been executed is called its linearization point.

Many distributed algorithms have been proposed, which build a read/write
register on top of a message-passing system, be it failure-free or
failure-prone. In the failure-prone case, the addressed failure models
are the process crash failure model, and the Byzantine process
failure model (see textbooks, e.g., ~\cite{AW04,L96,R10,R13}).
When considering process crash failures (the one considered in this
paper\footnote{For Byzantine failures, see for example~\cite{MPRJ17}.}),
the most famous of these algorithms was proposed by H. Attiya, A. Bar-Noy,
and D. Dolev in~\cite{ABD95}. This algorithm, usually called ABD
according to the names of its authors, considers an $n$-process asynchronous
system in which up to $t < n/2$ processes may crash.
As  $t < n/2$ is an upper bound of the number
of process crashes which can be tolerated (see~\cite{ABD95}),
this algorithm is $t$-resilient optimal. Its instances
implementing SWMR or MWMR atomic read/write registers rely on (a)
quorums~\cite{V12}, and (b) a classical broadcast/reply communication pattern.
This communication pattern is used twice in a read operation, and 
once (twice) in a write operation for an SWMR (MWMR) atomic 
read/write register. 

Other algorithms --each with its own properties-- implementing atomic
read/write registers on top of crash-prone asynchronous message-passing
systems can be found in the literature (\cite{A00,DGLV10,HNS16,MR16-podc} to
cite a few; see also the analytic presentation given in~\cite{R08}). 

\paragraph{From registers to snapshot objects}
The snapshot  object was introduced in~\cite{AADGMS93,A94}.
A snapshot object is an array $\REG[1..m]$ of atomic read/write registers
which provides the processes with two operations,
denoted $\wwrite()$ and $\snapshot()$.
If the base registers are SWMR  
the snapshot is called SWMR snapshot (and we have then $m=n$). In this case,
the invocation of $\wwrite(v)$ by a process $p_i$ assigns $v$ to $\REG[i]$, 
and the invocation of $\snapshot()$ by a process $p_i$ returns the
value of the full array as if the operation had been executed
instantaneously.
If the base registers are MWMR, the snapshot is called MWMR snapshot.
The invocation  of $\wwrite(r,v)$, where $1\leq r\leq m$,
by a process $p_i$ assigns $v$ to $\REG[r]$,
and  $\snapshot()$ is defined as before.
Said another way, the operations $\wwrite()$ and $\snapshot()$ are atomic,
i.e., in any execution of an SWMR (or MWMR) snapshot object, its operations 
$\wwrite()$ and $\snapshot()$ are linearizable.

Implementations of both  SWMR and  MWMR snapshot objects on top of
read/write atomic registers have been proposed
(e.g.,~\cite{AADGMS93,A94,IR12,ICMT94}). 
The ``hardness'' to build snapshot objects in read/write systems and
associated lower bounds are presented in the  survey~\cite{E05}.
The best algorithm known to implement  an SWMR snapshot requires 
$O(n \log n)$ read/write on the base SWMR registers  for both 
the $\wwrite()$ and $\snapshot()$ operations~\cite{AR98}.
As far as MWMR snapshot objects are concerned, there are implementations 
where each operation has an $O(n)$ cost\footnote{Snapshot objects built
  in read/write  models enriched with operations such as Compare\&Swap,
  or LL/SC, have also been considered, e.g.,\cite{J05,IR12}. Here
  we are interested in pure read/write models.}.

As far as the construction of an SWMR (or MWMR) snapshot object in 
crash-prone asynchronous message-passing systems where $t<n/2$ is
concerned, it is possible to stack two constructions: first 
an algorithm implementing SWMR (or MWMR) atomic read/write registers
(such as ABD), and, on top of it, an algorithm implementing an
SWMR (or MWMR) snapshot object. This stacking approach provides
objects whose operation cost is $O(n^2 \log n)$ messages for SWMR
snapshot, and $O(n^2)$ messages for MWMR snapshot.  An algorithm
based on the same communication pattern as ABD, which builds an atomic
SWMR snapshot object ``directly'' (i.e., without stacking algorithms)
was recently presented in~\cite{DFRR16} (the aim of this algorithm is to
perform better that the stacking approach in concurrency-free executions).

\paragraph{Another look at the implementation of read/write registers
and snapshot objects}
In sequential computing, there are ``natural'' pairings
linking data structures and control structures. The most
simple examples are the pair ``array and {\bf for} loop'', and
the pair ``tree and recursion''.

When we look at the implementation of a causal read/write
register~\cite{ANBHK95} on top of a (crash-free or crash-prone)
message-passing system, the causal message delivery broadcast
abstraction~\cite{BJ87,RST91} is the appropriate communication abstraction.
Namely, given this abstraction for free, 
the algorithms implementing the read and write operations build on top of it,
become very simple,  need only a few lines, and are easy to understand and to
prove correct. Of course, this is due to the fact that the  causal  broadcast
abstraction captures and abstracts the causality relation needed to
implement a causal read/write  register. Similarly, total order broadcast is the
communication abstraction associated with the consensus object~\cite{CT96}. 
This is summarized in Table~\ref{table1}.

\begin{table}[ht]
\begin{center}
\renewcommand{\baselinestretch}{1}
\small
\begin{tabular}{|c|c|}
\hline
Concurrent object &   Communication abstraction \\
\hline\hline
Causal read/write  registers &   Causal message delivery~\cite{BJ87,RST91} \\
\hline
Consensus~ &   Total order broadcast~\cite{CT96}\\
\hline
Snapshot object (and R/W register) &   \SCD-broadcast~(This paper)\\

\hline 
\end{tabular}
\end{center}
\caption{Associating objects and communication abstractions
                                              in a wait-free model}
\label{table1} 
\end{table}

As already said, all the algorithms we know which implement atomic
read/write registers, and (by stacking transitivity or directly)
SWMR or MWMR snapshots objects, on top of
crash-prone asynchronous message-passing systems, are based on a
broadcast/reply pattern plus the use of intersecting quorums.  Hence,
the following question naturally arises: Is this approach the ``only'' way
to implement a snapshot object (or an atomic register), or
is there a {\it specific
  communication abstraction which captures the essence and simplifies
  the implementation of snapshot objects (and atomic read/write registers)}?

\paragraph{Content of the paper}
Informatics in general (and  distributed computing in particular)
is a science of abstractions, and this paper is {\it distributed programming
abstraction}-oriented. It strives to address a
``desired level of abstraction and generality -- one that is broad
enough to encompass interesting new situations yet specific
enough to address the crucial issues'' as expressed in~\cite{FM03}. 
More precisely, it  answers the previous question in a positive way. 
To this end, it presents a simple broadcast abstraction which matches --and
therefore captures the essence of-- snapshot objects (and atomic read/write
registers). We call it {\it Set-Constrained Delivery Broadcast} (in
short \SCD-broadcast).  Given this communication abstraction, it is
possible to quorum-free build snapshot objects, and vice versa.
Hence,  similarly to consensus and total order broadcast, \SCD-broadcast
and snapshot objects have the same computational power (Table~\ref{table1}).

The \SCD-broadcast communication abstraction allows a process to
broadcast messages, and to deliver sets of messages (instead of single
messages) in such a way that, if a process $p_i$ delivers a message
set\footnote{In the rest of the paper, the identifiers starting
  with''{\it ms}'' denote message sets.} $ms$ containing a
message $m$, and later delivers a message set $ms'$ containing a
message $m'$, then no process $p_j$ can deliver first a set containing
$m'$ and later another set containing $m$.  Let us notice that $p_j$
is not prevented from delivering $m$ and $m'$ in the same set.


The implementation of an instance of \SCD-broadcast costs
$O(n^2)$ messages. It follows that  the cost  of a snapshot operation
(or a read/write register operation) on top of a message-passing
asynchronous system, where any minority of processes may crash, is
also $O(n^2)$ for both SWMR and  MWMR snapshot objects
(i.e., better than the stacking approach for SWMR snapshot objects).
Additionally, be the snapshot objects that are  built SWMR or MWMR, 
their implementation differ only in the fact that their underling
read/write registers are SWMR or MWMR. This provides us with a
noteworthy   genericity-related design simplicity. 

Of course, there is rarely something for free.  The algorithms
implementing the snapshot and write operations are simple because the
\SCD-broadcast abstraction hides enough ``implementation details'' and
provides consequently a high level abstraction (much higher than the
simple broadcast used in ABD-like algorithms). Its main interest lies
in its capture of the {\it high level message communication abstraction}
that, despite asynchrony and
process failures, allows simple message-passing implementations of shared
memory objects such as snapshot objects and atomic read/write registers.

\paragraph{Roadmap}
 The paper is composed of~\ref{sec:conclusion} sections.
Section~\ref{sec:models} presents the two base computation models
concerned in this paper, 
(read/write and message-passing). Section~\ref{sec:scd-broadcast}
presents the \SCD-broadcast communication abstraction.  Then,
Section~\ref{sec:from-SCD-to-snapshot} presents a simple algorithm
which implements a snapshot object on top of an asynchronous system
enriched with \SCD-broadcast, in which any number of processes may
crash.  Section~\ref{sec:from-snapshot-to-SCD} addresses the other
direction, namely, it presents an algorithm building the
\SCD-broadcast abstraction on top of an asynchronous system enriched
with snapshot objects and where any number of processes may crash.
Section~\ref{sec:conclusion} concludes the paper.
A noteworthy feature of the algorithms that are presented
lies in their simplicity, which is a first class property. 

Appendix~\ref{sec:algo-for-SC} describes an implementation
of \SCD-broadcast suited to  asynchronous message-passing systems where any
minority of processes may crash.  Hence, being implementable in the
weakest\footnote{\label{footnote1} From the point of view of the
  maximal number of process crashes that can be tolerated, assuming
  failures are independent.}  message-passing system model in which a
read/write register can be built, \SCD-broadcast is not ``yet another
oracle'' which makes things simpler to understand but cannot be
implemented.  Appendix~\ref{sec:from-SCD-to-SWMR-MWMR}
presents simplified \SCD-based algorithms which build atomic and 
sequentially consistent read/write registers.

\section{Basic Computation Models}
\label{sec:models}
This section presents two basic computation models.
In both cases, the process  model  is the same.

\subsection{Processes}
The computing model is composed of a set of $n$ asynchronous
sequential processes, denoted $p_1$, ..., $p_n$. ``Asynchronous'' means
that each process proceeds at its own speed, which can be arbitrary
and always remains  unknown to the other processes.  

A process may halt prematurely (crash failure), but it executes 
its local algorithm correctly until its possible crash. The model parameter $t$
denotes the maximal number of processes that may crash in a  run.
A process that crashes in a run is said to be {\it faulty}. Otherwise, 
it is {\it non-faulty}. Hence a faulty process behaves as a non-faulty process
until it crashes.

\subsection{Basic crash-prone asynchronous shared memory model}

\paragraph{Atomic read/write register}
The notion of an atomic read/write register has been formalized
in~\cite{L86,M86}. An MWMR {\it atomic} 
register (say $\REG$) is a concurrent object which provides each 
process  with an operation denoted $\REG.{\sf write}()$, 
and an operation denoted $\REG.{\sf read}()$.  
When a process invokes $\REG.{\sf write}(v)$ it defines
$v$ as being the new value of $\REG$.  An MWMR atomic register  is
defined by the following set of properties. 
\begin{itemize}
\item Liveness.
An invocation of an operation by a non-faulty process terminates.
\item Consistency (safety).  All the operations invoked by the
processes, except possibly --for each faulty process-- the last
operation it invoked, appear as if they have been executed
sequentially and this sequence of operations is such that:
\begin{itemize}
\item each read returns the value written by the closest write that precedes
it (or the initial value of $\REG$ if there is no preceding write), 
\item if an operation $op1$ terminates before an operation $op2$ starts, then 
 $op1$ appears before $op2$ in the sequence.
\end{itemize}
\end{itemize}
This set of properties states that, from an external observer point of
view, the read/write register appears as if it is accessed sequentially 
by the processes, and this sequence (a) respects the real-time access order,
and (b) belongs to the sequential specification of a register.

\paragraph{Notation}
The previous computation model is denoted ${\CARW}_{n,t}[\emptyset]$
(${\cal C}$rash ${\cal A}$synchronous ${\cal R}$ead-${\cal W}$rite).
This basic read/write model is also called {\it wait-free} read/write model. 
The symbol $\emptyset$ means there is no specific
constraint on $t$, which is equivalent to $t<n$, as it is always
assumed that not all processes crash.

\paragraph{Snapshot object}
This object was defined in the introduction. As we have seen,
snapshot objects can be built in ${\CARW}_{n,t}[\emptyset]$.
As we have seen there are two types of snapshot objects.
SWMR snapshot objects (whose base registers are SWMR), and
MWMR snapshot objects (whose base registers are MWMR).
In the following we consider MWMR snapshot objects, but the algorithms
can  be trivially adapted to work with  SWMR snapshot objects. 

 ${\CARW}_{n,t}[\emptyset]$ enriched with snapshot objects is
denoted  ${\CARW}_{n,t}[\mbox{snapshot]}$.
As a snapshot object can be built in  ${\CARW}_{n,t}[\emptyset]$
this model has the same computational power as ${\CARW}_{n,t}[\emptyset]$.
It only offers a higher abstraction level.

\subsection{Basic crash-prone asynchronous message-passing model}

\paragraph{Communication}
Each pair of processes communicate by sending and receiving messages
through two uni-directional channels, one in each direction. Hence,
the communication network is a complete network: any process $p_i$ can
directly send a message to any process $p_j$ (including itself).
A process $p_i$ invokes the operation ``${\sf send}$ {\sc type}($m$)
${\sf to}$ $p_j$'' to send to $p_j$ the message $m$, whose type is
{\sc type}.  The operation ``${\sf receive}$ {\sc type}() ${\sf from}$
$p_j$'' allows $p_i$ to receive from $p_j$ a message whose type is
{\sc type}.

Each channel is reliable (no loss, corruption, nor creation of messages),
not necessarily first-in/first-out, and asynchronous (while the transit time 
of each message is finite, there is no upper bound on message transit times).

Let us notice that, due to process and message asynchrony, no process can
know if another process crashed or is only very slow.

\paragraph{Notation and   necessary and sufficient condition}
This computation model is denoted ${\CAMP}_{n,t}[\emptyset]$
(${\cal C}$rash ${\cal A}$synchronous ${\cal M}$essage-${\cal  P}$assing).

The constraint $(t<n/2)$ is a necessary and sufficient condition to implement 
an atomic read/write register in ${\CAMP}_{n,t}[\emptyset]$~\cite{ABD95}. 
Hence, the model  ${\CAMP}_{n,t}[\emptyset]$ whose runs are constrained by 
$t<n/2$  is denoted ${\CAMP}_{n,t}[t<n/2]$.

\section{A Broadcast Abstraction: Set-Constrained Message Delivery}
\label{sec:scd-broadcast}

\paragraph{Definition}
The set-constrained broadcast abstraction (\SCD-broadcast)
provides the processes with
two operations, denoted $\scbroadcast()$ and  $\scdeliver()$.
The first operation takes a message to broadcast as input  parameter. 
The second one returns a non-empty set of messages to the process that
invoked it. 
Using a classical terminology, when a process invokes $\scbroadcast(m)$,
we say that it ``scd-broadcasts a message $m$''. Similarly, when it invokes 
$\scdeliver()$ and obtains a set of messages $ms$, we say that
it ``scd-delivers a set of messages $ms$''. By a slight abuse of language,
we also say that a process ``scd-delivers a  message $m$'' when it  delivers 
a message  $m\in ms$.

\SCD-broadcast is defined by the following
set of properties, where we assume --without loss of generality--
that all the messages that are scd-broadcast are different. 
\begin{itemize}
\item Validity.
  If a process scd-delivers a set containing a message $m$,
  then $m$ was scd-broadcast by some process.
\item Integrity.
  A message is scd-delivered at most once by each process.
\item MS-Ordering.
  If a process $p_i$ scd-delivers first a message $m$ belonging to a set
  $ms_i$ and later a message $m'$ belonging to a set $ms_i'\neq ms_i$, then 
  no process  scd-delivers first the message $m'$ in some
  scd-delivered set $ms'_j$ and later the message $m$ in some scd-delivered
  set $ms_j\neq ms'_j$. 
\item Termination-1.
  If a non-faulty process scd-broadcasts a message $m$,
  it terminates its scd-broadcast invocation and 
  scd-delivers a message set containing $m$.
\item Termination-2.
  If a non-faulty process scd-delivers a message $m$, every  non-faulty process
  scd-delivers a message set containing $m$.
\end{itemize}

Termination-1 and Termination-2 are classical liveness properties
(found for example in Uniform Reliable Broadcast). 
The other ones are safety properties.
Validity and Integrity are classical communication-related properties.
The first  states that there is neither  message creation nor message
corruption, while the second states that there is no message duplication.

The MS-Ordering property is new,
and characterizes \SCD-broadcast. It states that the contents of the sets of
messages  scd-delivered at any two processes are not totally independent:
the sequence of sets scd-delivered at a process $p_i$ and
the sequence of sets scd-delivered at a process $p_j$ must be mutually
consistent in the sense that a process $p_i$ cannot scd-deliver
first $m\in ms_i$ and later $m'\in ms_i'\neq ms_i$, while another process 
$p_j$  scd-delivers first $m'\in ms_j'$ and later $m\in ms_j\neq ms_j'$. 
Let us nevertheless observe that if $p_i$  scd-delivers
first $m\in ms_i$ and later $m'\in ms_i'$,
$p_j$ may scd-deliver $m$ and $m'$ in the same set of messages.

\paragraph{An example}
Let $m_1$, $m_2$, $m_3$,  $m_4$, $m_5$, $m_6$, $m_7$, $m_8$, ... be
messages that have been scd-broadcast by different processes.
The following  scd-deliveries of message sets by $p_1$, $p_2$ and $p_3$
respect the definition of \SCD-broadcast: 
\begin{itemize}
\item at $p_1$:
$\{m_1,m_2\}$, $\{m_3,m_4,m_5\}$, $\{m_6\}$, $\{m_7,m_8\}$.
\item at $p_2$:
$\{m_1\}$, $\{m_3,m_2\}$, $\{m_6,m_4,m_5\}$, $\{m_7\}$, $\{m_8\}$.
\item at $p_3$:
$\{m_3,m_1,m_2\}$, $\{m_6,m_4,m_5\}$, $\{m_7\}$, $\{m_8\}$.  
\end{itemize}
Differently, due to the scd-deliveries of the sets including $m_2$ and $m_3$, 
the following  scd-deliveries by $p_1$ and $p_2$ do not satisfy  the
MS-broadcast property: 
\begin{itemize}
\item at $p_1$:
$\{m_1,m_2\}$, $\{m_3,m_4,m_5\}$, ...  
\item at $p_2$:
$\{m_1,m_3\}$, $\{m_2\}$, ...
\end{itemize}

\paragraph{A containment property}
Let $ms_i^\ell$ be the  $\ell$-th message set  scd-delivered by $p_i$.
Hence, at some time, $p_i$ scd-delivered the sequence
of message sets $ms_i^1,~\cdots, ms_i^x$.
Let $\MS_i^x= ms_i^1\cup \cdots \cup ms_i^x$. 
The following  property follows directly from the 
MS-Ordering and Termination-2 properties: 
\begin{itemize}
\item Containment.
$\forall~i,j,x,y$:
$(\MS_i^x \subseteq \MS_j^y) \vee (\MS_j^y\subseteq  \MS_i^x)$.
\end{itemize}

\paragraph{Remark 1: Weakening SCD-broadcast}
If the messages in a message set are delivered one at a time, and
the MS-Ordering property is suppressed, \SCD-broadcast boils down to
{\it Reliable Broadcast}. 

\paragraph{Remark 2: On the partial order created by the  message sets}
The MS-Ordering  and Integrity properties establish
a partial order on the set of all the messages, defined as follows.
Let $\mapsto_i$ be the local message delivery order at a process
$p_i$ defined as follows:
$m \mapsto_i m'$ if $p_i$
scd-delivers the set containing $m$ before the set containing $m'$. 
As no message is scd-delivered twice, it is easy to see that 
$ \mapsto_i$ is a  partial order (locally know by $p_i$). 
The reader can check that there is a total order (which remains
unknown to the processes)  on the whole set of messages,
that complies with the  partial order $\cup_{1\leq i\leq n}\mapsto_i$. 
This is where \SCD-broadcast can be seen as a weakening
of total order broadcast.

\Xomit{
Let $\rightarrow_{msg}~\stackrel{\mathit{def}}{=}
\cup_{1\leq i \leq n}\rightarrow_i$.
Due to the MS-Ordering  property, this  relation  is partial order
on the set of all the  messages  that have been scd-broadcast.
It is  global in the sense that
it involves all the local relations $\rightarrow_i$, and consequently
cannot be fully  known by each process taken individually. 
}

\section{From \SCD-broadcast to an MWMR Snapshot Object}
\label{sec:from-SCD-to-snapshot}
Let  $\CAMP_{n,t}[\mbox{\SCD-broadcast}]$ denote
$\CAMP_{n,t}[\emptyset]$ enriched with the  \SCD-broadcast abstraction.
Hence, this abstraction is given for free. 
This section presents and proves correct a simple algorithm building an 
MWMR snapshot object  on top of $\CAMP_{n,t}[\mbox{\SCD-broadcast}]$.
The same algorithm with very few  simple modifications can be used to
build SWMR or MWMR atomic registers in  $\CAMP_{n,t}[\mbox{\SCD-broadcast}]$
(see Appendix~\ref{sec:from-SCD-to-SWMR-MWMR}).

\subsection{Building an MWMR snapshot object  on top
  of $\CAMP_{n,t}[\mbox{\SCD-broadcast}]$}
Let $\REG[1..m]$ denote the MWMR snapshot object  that is built.


\paragraph{Local representation of  $\REG$ at a process $p_i$}
At each register $p_i$,  $\REG[1..m]$ is represented by three local variables
$reg_i[1..m]$ (data part), plus  $tsa_i[1..m]$ and $done_i$ (control part). 
\begin{itemize}
\item $\done_i$ is a Boolean variable. 
\item $\reg_i[1..m] $ contains the current value of  $\REG[1..m]$,
  as known by $p_i$.
\item $tsa_i[1..m]$ is an array of timestamps associated with the values
  stored in $\reg_i[1..m]$.
A timestamp is a pair made of a local clock value and a process identity. 
Its initial value is $\langle 0,-\rangle$.
The fields associated with $tsa_i[r]$ are denoted
 $\langle tsa_i[r].date ,tsa_i[r].proc \rangle$. 
\end{itemize}

\paragraph{Timestamp-based order relation}
We consider the classical lexicographical total order relation
on timestamps, denoted  $<_{ts}$. Let $ts1= \langle h1,i1\rangle$ and
$ts2= \langle h2,i2\rangle$. We have 
$ ts1 \lexts ts2 \stackrel{\mathit{def}}{=}
(h1<h2) \vee ((h1=h2)\wedge (i1<i2)).$

\paragraph{Algorithm~\ref{algo:snapshot-from-SCD}: snapshot operation}
(Lines~\ref{Snap-from-SC-01}-\ref{Snap-from-SC-04}) When $p_i$ invokes
$\REG.\snapshot()$, it first sets $\done_i$ to $\ffalse$, and invokes
$\scbroadcast$ \SYNC$(i)$.  \SYNC() is a synchronization message,
whose aim is to entail the refreshment of the value of $\reg_i[1..m]$
(lines~\ref{Snap-from-SC-11}-\ref{Snap-from-SC-17}) which occurs before
the setting of $\done_i$ to $\ttrue$ (line~\ref{Snap-from-SC-18}).
When this happens, $p_i$ returns the value of its local variable
$\reg_i[1..m]$ and terminates its snapshot invocation.

\begin{algorithm}[h!]
\centering{\fbox{
\begin{minipage}[t]{150mm}
\footnotesize 
\renewcommand{\baselinestretch}{2.5}
\resetline
\begin{tabbing}
aaa\=aa\=aaa\=aaaaaa\=\kill

{\bf operation} $\snapshot()$ {\bf is}\\

\line{Snap-from-SC-01}  \>\> $\done_i \leftarrow \ffalse$;\\
\line{Snap-from-SC-02}  \>\> $\scbroadcast$ \SYNC $(i)$; \\
\line{Snap-from-SC-03}  \>\>  $\wait (\done_i)$;\\
\line{Snap-from-SC-04}  \>\> $\return (\reg_i[1..m])$. \\~\\

{\bf operation} $\wwrite(r, v)$ {\bf is}\\

\line{Snap-from-SC-05}  \>\> $\done_i \leftarrow \ffalse$;\\
\line{Snap-from-SC-06}  \>\> $\scbroadcast$ \SYNC $(i)$; \\
\line{Snap-from-SC-07}  \>\>  $\wait (\done_i)$;\\

\line{Snap-from-SC-08}  \>\> $\done_i \leftarrow \ffalse$;\\
\line{Snap-from-SC-09}  \>\>
        $\scbroadcast$ \WRITE $(r,v, \langle \tsa_i[r].date +1, i \rangle)$; \\
\line{Snap-from-SC-10}  \>\> $\wait (\done_i)$.~\\~\\

{\bf when the message set} \= 
$\{$~\WRITE$(r_{j_1}, v_{j_1}, \langle date_{j_1}, j_1\rangle),~\cdots,$
     \WRITE$(r_{j_x}, v_{j_x}, \langle date_{j_x}, j_x\rangle),$ \\

\>\>$~~~~~~~~~~~~~~~~~~~~~~~~~~~~~$
   \SYNC$(j_{x+1}),~\cdots,$ \SYNC$(j_{y})~\}$ {\bf is scd-delivered do}\\

\line{Snap-from-SC-11} 
\>\> {\bf for} \= {\bf  each} 
   $r$ such that  \WRITE$(r,-,-)$ $\in$  scd-delivered message set {\bf do}\\

\line{Snap-from-SC-12} 
\>\>\> {\bf let} $\langle date, writer\rangle$ be the greatest timestamp
in the messages  \WRITE$(r,-,-)$;\\

\line{Snap-from-SC-13} 
\>\>\> {\bf if} \=  $(\tsa_i[r] \lexts \langle date,writer \rangle $) \\

\line{Snap-from-SC-14}
\>\>\>\>  {\bf then} \=

   {\bf let} $v$ the value in  \WRITE$(r,-,\langle date,writer \rangle)$;\\

\line{Snap-from-SC-15}
 \>\>\>\>\>
 $reg_i[r]\leftarrow v$; $\tsa_i[r]\leftarrow  \langle date,writer \rangle$\\

\line{Snap-from-SC-16}
\>\>\> {\bf end if}\\

\line{Snap-from-SC-17}
\>\> {\bf end for};\\

\line{Snap-from-SC-18}  \>\>
              {\bf if} $\exists \ell~:~j_\ell=i$ 
                {\bf then} $\done_i \leftarrow \ttrue$ {\bf end if}. 

\end{tabbing}
\end{minipage}
}
  \caption{Construction of an MWMR  snapshot object 
           $\CAMP_{n,t}[\mbox{\SCD-broadcast}]$ (code for $p_i$)}
\label{algo:snapshot-from-SCD}
}
\end{algorithm}

\paragraph{Algorithm~\ref{algo:snapshot-from-SCD}: write operation}
(Lines~\ref{Snap-from-SC-05}-\ref{Snap-from-SC-10})
When a process $p_i$ wants to assign a value $v$ to $\REG[r]$, 
it invokes $\REG.\wwrite(r,v)$.
This operation is made up of two parts.
First $p_i$ executes a re-synchronization
(lines~\ref{Snap-from-SC-05}-\ref{Snap-from-SC-07}, exactly 
as in the snapshot operation) whose side
effect is here to provide $p_i$ with an up-to-date value of $\tsa_i[r].date$.
In the second part, $p_i$ associates the timestamp $\langle
\tsa_i[r].date+1,i\rangle$ with $v$, and invokes $\scbroadcast$
\WRITE$(r, v, \langle \tsa_i[r].date+1,i\rangle)$
(line~\ref{Snap-from-SC-09}).  In addition to informing the other
processes on its write of $\REG[r]$, this message \WRITE$()$ acts as a
re-synchronization message, exactly as a message \SYNC$(i)$.  When
this synchronization terminates (i.e., when the Boolean $\done_i$ is
set to ${\ttrue}$), $p_i$ returns from the write operation
(line~\ref{Snap-from-SC-10}).

\paragraph{Algorithm~\ref{algo:snapshot-from-SCD}:
  scd-delivery of a set of messages}
When $p_i$ scd-delivers a message set, namely, \\
\centerline{
$\{$~\WRITE$(r_{j_1}, v_{j_1}, \langle date_{j_1}, j_1\rangle),~\cdots,$
     \WRITE$(r_{j_x}, v_{j_x}, \langle date_{j_x}, j_x\rangle),$ 
      \SYNC$(j_{x+1}),~\cdots,$ \SYNC$(j_{y})~\}$}\\ 
it first looks if  there are messages \WRITE$()$.
If it is the case, for each register $\REG[r]$ for which there are 
messages \WRITE$(r,-,-)$   (line~\ref{Snap-from-SC-11}),
$p_i$ computes the maximal timestamp carried
by these messages (line~\ref{Snap-from-SC-12}), and  updates accordingly
its local representation of $\REG[r]$
(lines~\ref{Snap-from-SC-13}-\ref{Snap-from-SC-15}).
Finally, if $p_i$ is the sender of one of these messages
(\WRITE$()$ or \SYNC$()$), 
$done_i$ is set to $\ttrue$, which terminates $p_i$'s re-synchronization
(line~\ref{Snap-from-SC-18}).

\paragraph{Message cost}
An invocation of  $\snapshot()$ involves one
invocation of $\scbroadcast()$, and an  invocation of
$\wwrite()$ involves two such invocations. 
It is shown in Appendix~\ref{sec:algo-for-SC} that,
in a message-passing system, $\scbroadcast()$ costs $O(n^2)$ protocol
messages. It follows that, in such systems,  the message cost of both
operations of a snapshot object  is $O(n^2)$. 
(This remains true for SWMR snapshot objects, see
Appendix~\ref{sec:from-SCD-to-SWMR-MWMR}.)

\subsection{Proof of Algorithm~\ref{algo:snapshot-from-SCD}}
\label{sec:proof-snapshot}
As they are implicitly used in the  proofs that follow, let us recall
the properties of the \SCD-broadcast abstraction. The non-faulty processes
scd-deliver the same messages (exactly one each), and
each of them was scd-broadcast. As a faulty process behaves correctly
until it crashes, it scd-delivers a subset of the messages 
scd-delivered by the non-faulty processes. 

Without loss of generality, we assume that there is an initial write
operation issued by a non-faulty process.
Moreover, if  a process crashes in a snapshot operation, its snapshot is not
considered;  if  a process crashes in a write operation, its
write is considered only if the message \WRITE() it sent at
line~\ref{Snap-from-SC-09} is scd-delivered to at least one non-faulty process
(and by the Termination-2 property, at least to all non-faulty processes). 
Let us notice that  a message  \SYNC() scd-broadcast by a process $p_i$
does not modify the local variables of  the other processes.

\section{Proof of Lemmas for Theorem~\ref{theorem:proof-snapshot}}
\label{sec:appendix-proof-theorem-1}

\begin{lemma}
\label{lemma:snapshot-liveness}
If a non-faulty process invokes an operation, it returns from its invocation. 
\end{lemma}

\begin{proofL}
  Let $p_i$ be a non-faulty process that invokes a read or write  operation.
  By the Termination-1 property of \SCD-broadcast, it 
  eventually receives a message set containing the message
  \SYNC() or \WRITE() it sends at 
  line \ref{Snap-from-SC-02}, \ref{Snap-from-SC-06} or \ref{Snap-from-SC-09}.
  As all the statements associated with the scd-delivery of a message set
  (lines~\ref{Snap-from-SC-11}-\ref{Snap-from-SC-18}) terminate, 
  it follows that  the synchronization Boolean $\done_i$   is eventually
  set to $\ttrue$. Consequently,  $p_i$ returns from the invocation of its
  operation.
\renewcommand{\toto}{lemma:snapshot-liveness}
\end{proofL}

\paragraph{Extension of the relation $\lexts$}
The relation $\lexts$ is extended to  
a partial order on arrays of timestamps, denoted $ \le_{\tsa}$,
defined as follows: 
$tsa1[1..m] \le_{\tsa} tsa2[1..m] \stackrel{\mathit{def}}{=}
\forall r: (\tsa1[r] = \tsa2[r] \lor \tsa1[r] \lexts\ tsa2[r])$.
Moreover,  $\tsa1[1..m] <_{\tsa} \tsa2[1..m] \stackrel{\mathit{def}}{=}
(\tsa1[1..m] \le_{tsa} \tsa2[1..m])\wedge (\tsa1[1..m] \neq tsa2[1..m])$.
\paragraph{Definition}
Let $\TSA_i$ be the set of the array values  taken by $ts_i[1..m]$ at
line~\ref{Snap-from-SC-18} (end of the processing of  a message set
by process $p_i$). Let $\TSA=\cup_{1\leq i\leq n} \TSA_i$. 

\begin{lemma}
\label{lemma:clock-ordering}
The order $\le_{\tsa}$ is total on $\TSA$.
\end{lemma}

\begin{proofL}
  Let us first observe that, for any $i$,  all values in $\TSA_i$
  are totally ordered (this comes from $ts_i[1..m]$ whose entries can
  only increase,  lines~\ref{Snap-from-SC-13} and~\ref{Snap-from-SC-15}).  
  Hence, let $\tsa1[1..m]$ be an array value
  of $\TSA_i$, and  $\tsa2[1..m]$ an array value of $\TSA_j$, where $i\neq j$.

  Let us assume, by contradiction, that $\lnot (\tsa1 \le_{\tsa}
  \tsa2)$ and $\lnot (\tsa2 \le_{\tsa} \tsa1)$.  As $\lnot (\tsa1
  \le_{\tsa} \tsa2)$, there is a registers $r$ such that $\tsa2[r] <\tsa1[r]$.
  According to lines~\ref{Snap-from-SC-13}
  and~\ref{Snap-from-SC-15}, there is a message \WRITE$(r,-,  \tsa1[r])$
  received by $p_i$ when $\tsa_i = \tsa1$ and not received  
  by $p_j$ when $\tsa_j = \tsa2$ (because $\tsa2[r] <\tsa1[r]$).
  Similarly, there is a message
  \WRITE$(r',-, \tsa2[r'])$ received by $p_j$ when $\tsa_j = \tsa2$
  and not received by $p_i$ when $\tsa_i = \tsa1$.
  This situation contradicts the MS-Ordering property, from which we conclude
  that either $\tsa1 \le_{\tsa} \tsa2$ or $\tsa2 \le_{\tsa} \tsa1$. 
  \renewcommand{\toto}{lemma:clock-ordering} 
\end{proofL}

\paragraph{Definitions}
Let us associate a timestamp $ts(\wwrite(r, v))$ with each write
operation as follows. Let $p_i$ be the invoking process; $ts(\wwrite(r,
v))$ is the timestamp of $v$ as defined by $p_i$ at
line~\ref{Snap-from-SC-09}, i.e., $\langle \tsa_i[r].date+1, i\rangle$.

Let $\op1$ and $\op2$ be any two operations. The relation
$\prec$ on the whole set of operations is defined as follows:
 $\op1 \prec \op2$ if $\op1$ terminated before $\op2$ started. 
It is easy to see that $\prec$ is a real-time-compliant partial order 
on all the operations.

\begin{lemma}
\label{lemma:snapshot-write-ordering}
No two distinct write operations on the same register $\wwrite1(r, v)$
and $\wwrite2(r, w)$ have the same timestamp, and $(\wwrite1(r, v)
\prec\wwrite2(r, w))$ $\Rightarrow$ $(ts(\wwrite1) \lexts ts(\wwrite2))$.
\end{lemma}

\begin{proofL}
Let $\langle date1,i\rangle$ and $\langle date2,j\rangle$ be the
timestamp of $\wwrite1(r, v)$ and $\wwrite2(r, w)$, respectively.  If
$i\neq j$, $\wwrite1(r, v)$ and $\wwrite2(r, w)$ have been produced by
different processes, and their timestamp differ at least in their
process identity.

So, let us consider that the operations have been issued by the same
process $p_i$, with $\wwrite1(r, v)$ first.  As $\wwrite1(r, v)$
precedes $\wwrite2(r, w)$, $p_i$ first invoked  $\scbroadcast$
\WRITE$(r, v,\langle date1,i\rangle)$ (line~\ref{Snap-from-SC-09}) and
later \WRITE$(r, w,\langle date2,i\rangle)$. It follows that these
\SCD-broadcast invocations are separated by a local reset of the Boolean
$done_i$ at line~\ref{Snap-from-SC-16}.  Moreover, before the reset of
$\done_i$ due to the scd-delivery of
the message $\{ \cdots, $\WRITE$(r, v,\langle date1,i\rangle),\cdots\}$,
we have $\tsa_i[r].date_i\geq date1$
(lines~\ref{Snap-from-SC-12}-\ref{Snap-from-SC-16}).  Hence, we have
$\tsa_i[r].date\geq date1$ before the reset of $\done_i$
(line~\ref{Snap-from-SC-18}).  Then, due to the ``$+1$'' at
line~\ref{Snap-from-SC-09}, \WRITE $(r, w,\langle date2,i\rangle)$ is
such that $date2>date1$, which concludes the proof of the first part
of the lemma. 

Let us now consider that  $\wwrite1(r, v) \prec\wwrite2(r, w)$. If
$\wwrite1(r, v)$ and $\wwrite2(r, w)$ have been produced by the same process
we have $date1 < date2$ from the previous reasoning.
So let us assume that they have been produced by different processes
$p_i$ and $p_j$.
Before  terminating $\wwrite1(r, v)$ (when the Boolean $\done_i$ is set
$\ttrue$  at line~\ref{Snap-from-SC-18}), $p_i$  received a message set
$ms1_i$ containing the message \WRITE$(r, v,\langle date1,i\rangle)$. 
When $p_j$ executes  $\wwrite2(r, w)$, it first  invokes
$\scbroadcast$ \SYNC$(j)$ at line~\ref{Snap-from-SC-06}. 
Because $\wwrite1(r, v)$ terminated before  $\wwrite2(r, w)$ started, 
this message \SYNC$(j)$ cannot belong to $ms1_i$.

Due to Integrity and Termination-2 of \SCD-broadcast, $p_j$ eventually
scd-delivers exactly one  message set $ms1_j$ containing
\WRITE$(r, v,\langle date1,i\rangle)$. Moreover, it also
scd-delivers  exactly one  message set $ms2_j$ containing
its own message \SYNC$(j)$.
On the the other side, $p_i$ scd-delivers exactly one  message
set $ms2_i$ containing the message \SYNC$(j)$.
It follows from the MS-Ordering property that, if  $ms2_j\neq ms1_j$, 
$p_j$ cannot scd-deliver $ms2_j$  before  $ms1_j$.
Then, whatever the case ($ms1_j=ms2_j$ or  $ms1_j$ is scd-delivered at
$p_j$ before $ms2_j$), it follows from the fact that the messages \WRITE$()$
are processed (lines~\ref{Snap-from-SC-11}-\ref{Snap-from-SC-17})
before the messages \SYNC$(j)$ (line~\ref{Snap-from-SC-18}), that we have
$\tsa_j[r]\geq \langle date1,i\rangle$ when $\done_j$ is set to $\ttrue$. 
It then follows from line~\ref{Snap-from-SC-09} that $date2>date1$,
which concludes the proof of the lemma. 
\renewcommand{\toto}{lemma:snapshot-write-ordering}
\end{proofL}

\paragraph{Associating timestamp arrays with operations}
Let us associate a timestamp array $\tsa(\op)[1..m]$ with each operation
$\op()$ as follows.
\begin{itemize}
\item
  Case $\op()=\snapshot()$. Let $p_i$ be the invoking process;
  $\tsa(\op)$ is the value of $\tsa_i[1..m]$ when $p_i$ returns from the
  snapshot operation (line~\ref{Snap-from-SC-04}).  
\item Case $\op()=\wwrite(r,v)$. 
  Let $\mmin_\tsa(\{A\})$, where $A$ is a set of array values,      
  denote the smallest array value of $A$ according to $<_\tsa$. 
  Let $\tsa(\op)\stackrel{\mathit{def}}{=}   
  \mmin_{\tsa}(\{\tsa[1..m] \in \TSA
                    \mbox{ such  that } ts(\op) \leq_{ts} \tsa[r]\})$.    
   Hence,  $\tsa(\op)$ is the first $\tsa[1..m]$ of $\TSA$, that reports
   the operation $\op()=\wwrite(r,v)$.        
\end{itemize}

\begin{lemma}
\label{lemma:snapshot-clock-ordering}
Let $\op$ and $\op'$ be two distinct operations such that $\op
\prec\op'$. We have $\tsa(\op) \le_{\tsa} \tsa(\op')$. Moreover, if $\op'$
is a write operation, we have $\tsa(\op) <_{\tsa} \tsa(\op')$.
\end{lemma}

\begin{proofL}
  Let $p_i$ and $p_j$ be the processes that performed $\op$ and 
  $\op'$, respectively.
  Let \SYNC$_j$ be the \SYNC$(j)$ message sent by $p_j$ (at
  line~\ref{Snap-from-SC-02} or~\ref{Snap-from-SC-06}) during the
  execution of $\op'$.  Let $\mathit{term\_\tsa_i}$ be the
  value of $\tsa_i[1..m]$ when $\op$ terminates
  (line~\ref{Snap-from-SC-04} or~\ref{Snap-from-SC-10}), and
  $\mathit{sync\_\tsa_j}$
  the value of $\tsa_j[1..m]$ when $\mathit{done_j}$ becomes true for the first
  time after $p_j$ sent \SYNC$_j$ (line~\ref{Snap-from-SC-03}
  or~\ref{Snap-from-SC-07}). Let us notice that $\mathit{term\_\tsa_i}$
  and  $\mathit{sync\_\tsa_j}$ are elements of the set $\TSA$.

  According to lines~\ref{Snap-from-SC-13} and~\ref{Snap-from-SC-15}, for
  all $r$, $\tsa_i[r]$ is the largest timestamp carried by
  a message \WRITE$(r,v,-)$
  received by $p_i$ in a message set before $\op$ terminates.  Let $m$ be a
  message such that there is a set $sm$ scd-delivered by $p_i$ before it
  terminated $\op$. As $p_j$ sent \SYNC$_j$ after $p_i$ terminated,
  $p_i$ did not receive any set containing \SYNC$_j$ before it
  terminated $\op$. By the properties Termination-2 and MS-Ordering, $p_j$
  received message $m$ in the same set as \SYNC$_j$ or in a message
  set $sm'$ received before the set containing \SYNC$_j$. Therefore,
  we have $\mathit{term\_\tsa_i} \le_{\tsa} \mathit{sync\_\tsa_j}$.
  
  If $\op$ is a snapshot operation, then $\tsa(\op) =
  \mathit{term\_\tsa_i}$. Otherwise, $\op()=\wwrite(r,v)$. As $p_i$ has
  to wait until it processes a set of messages including  its \WRITE() message
  (and executes line~\ref{Snap-from-SC-18}),
   we have $ts(\op) \lexts \mathit{term\_\tsa_i}[r]$.
   Finally, due to the fact that $\mathit{term\_\tsa_i}\in \TSA$ and
   Lemma~\ref{lemma:clock-ordering}, we have
   $\tsa(\op)\le_{\tsa}\mathit{term\_\tsa_i}$.

  If $\op'$ is a snapshot operation, then $\mathit{sync\_\tsa_j} = \tsa(\op')$
   (line~\ref{Snap-from-SC-04}).
  Otherwise, $\op()=\wwrite(r,v)$ and thanks to the $+1$ in
  line~\ref{Snap-from-SC-09},
  $\mathit{sync\_\tsa_j}[r]$ is strictly smaller than $\tsa(\op')[r]$
  which, due to Lemma~\ref{lemma:clock-ordering}, implies
  $\mathit{sync\_\tsa_j} <_{\tsa} \tsa(\op')$.

  It follows that, in all cases, we have
  $\tsa(\op) \le_{\tsa} \mathit{term\_\tsa_i} \le_{\tsa} \mathit{sync\_\tsa_j}
                                                \le_{\tsa} \tsa(\op')$
  and if $\op'$ is a write operation, we have $\tsa(\op) \le_{\tsa}
  \mathit{term\_\tsa_i} \le_{\tsa} \mathit{sync\_\tsa_j} <_{\tsa} \tsa(\op')$,
  which concludes the proof of the lemma.
  \renewcommand{\toto}{lemma:snapshot-clock-ordering}
\end{proofL}


The previous lemmas allow the operations to be linearized (i.e.,
totally ordered in an order compliant with both the sequential
specification of a register, and their real-time occurrence order)
according to a total order extension of the reflexive and transitive
closure of the $\rightarrow_{lin}$ relation defined thereafter.
\begin{definition}
  Let $\op, \op'$ be two operations. We define the $\rightarrow_{lin}$ relation
  by $\op \rightarrow_{lin} \op'$ if one of the following properties holds:
  \begin{itemize}
  \item
    $\op \prec \op'$,
  \item
    $tsa(\op) <_{tsa} tsa(\op')$,
  \item
    $tsa(\op) = tsa(\op')$, $op$ is a write operation
    and $\op'$ is a snapshot operation,
  \item
    $tsa(\op) = tsa(\op')$, $\op$ and $\op'$ are two
    write operations on the same register and $ts(\op) \lexts ts(\op')$,
  \end{itemize}
\end{definition}

\begin{lemma}
 \label{lemma:snapshot-safety}
 The snapshot object built by Algorithm~{\em{\ref{algo:snapshot-from-SCD}}}
 is linearizable.
\end{lemma}

\begin{proofL}
  We recall the definition of the $\rightarrow_{lin}$ relation: $\op
  \rightarrow_{lin} \op'$ if one of the following properties holds:
  \begin{itemize}
  \item
    $\op \prec \op'$,
  \item
    $tsa(\op) <_{tsa} tsa(\op')$,
  \item
    $tsa(\op) = tsa(\op')$, $op$ is a write operation
    and $\op'$ is a snapshot operation,
  \item
    $tsa(\op) = tsa(\op')$, $\op$ and $\op'$ are two
    write operations on the same register and $ts(\op) \lexts ts(\op')$,
  \end{itemize}
  We define the $\rightarrow_{lin}^\star$ relation as the reflexive
  and transitive closure of the $\rightarrow_{lin}$ relation.

  Let us prove that the $\rightarrow_{lin}^\star$ relation is a
  partial order on all operations. Transitivity and reflexivity are
  given by construction.  Let us prove antisymmetry. Suppose there are
  $\op_0, \op_2, ..., \op_m$ such that $\op_0 = \op_m$ and $\op_i
  \rightarrow_{lin} \op_{i+1}$ for all $i<m$.  By
  Lemma~\ref{lemma:snapshot-clock-ordering}, for all $i<m$, we have
  $tsa(\op_i) \le_{tsa} tsa(\op_{i+1})$, and $tsa(\op_m) =
  tsa(\op_{0})$, so the timestamp array of all operations are the
  same.  Moreover, if $\op_i$ is a snapshot operation, then $\op_i
  \prec \op_{(i+1) \% m}$ is the only possible case
  ($\%$ stands for ``modulo'') , and by
  Lemma~\ref{lemma:snapshot-clock-ordering} again, $\op_{(i+1) \% m}$
  is a snapshot operation. Therefore, only two cases are possible.
  
  \begin{itemize}
  \item Let us suppose that all the $\op_i$ are snapshot operations
    and for all $i$, $\op_i \prec \op_{(i+1) \% m}$. As $\prec$ is a
    partial order relation, it is antisymmetric, so all the $\op_i$
    are the same operation.
  \item Otherwise, all the $\op_i$ are write operations. By
    Lemma~\ref{lemma:snapshot-clock-ordering}, for all $\op_i
    \not\prec \op_{(i+1) \% m}$.  The operations $\op_i$ and
    $\op_{i+1\% m}$ are ordered by the fourth point, so they are write
    operations on the same register and $ts(\op_i) \lexts
    ts(\op_{i+1\% m})$. By antisymmetry of the $\lexts$ relation, all
    the $\op_i$ have the same timestamp, so by
    Lemma~\ref{lemma:snapshot-write-ordering}, they are the same
    operation, which proves antisymmetry.
  \end{itemize}
  Let $\le_{lin}$ be a total order extension of
  $\rightarrow_{lin}^\star$. Relation $\le_{lin}$ is real-time
  compliant because $\rightarrow_{lin}^\star$ contains $\prec$.
  
  Let us consider a snapshot operation $\op$ and a register $r$ such
  that $tsa(\op)[r] = \langle date1, i \rangle$. According to
  line~\ref{Snap-from-SC-10}, it is associated to the value $v$ that
  is returned by $\rread1()$ for $r$, and comes from a
  \WRITE$(r,v,\langle date1, i \rangle)$ message sent by a write
  operation $\op_r = \wwrite(r,v)$.  By definition of $tsa(\op_r)$, we have
  $tsa(\op_r) \le_{tsa} tsa(\op)$ (Lemma~\ref{lemma:snapshot-clock-ordering}), 
  and therefore $\op_r \le_{lin}
  \op$. Moreover, for any different write operation $\op'_r$ on $r$,
  by Lemma~\ref{lemma:snapshot-write-ordering}, $ts(\op'_r) \neq
  ts(\op_r)$. If $ts(\op'_r) \lexts ts(\op_r)$, then $\op'_r \le_{lin}
  \op_r$. Otherwise, $tsa(\op) <_{tsa} tsa(\op'_r)$, and (due to the first item
  of the definition of  $\rightarrow_{lin}$) we have  $\op \le_{lin}  \op'_r$.
  In both cases, the value written by $\op_r$ is the last
  value written on $r$ before $\op$, according to $\le_{lin}$.
  \renewcommand{\toto}{lemma:snapshot-safety}
\end{proofL}

\begin{theorem}
\label{theorem:proof-snapshot}
Algorithm~{\em{\ref{algo:snapshot-from-SCD}}} builds an {\em MWMR}  snapshot
object in the system model  $\CAMP_{n,t}[\mbox{\em \SCD-broadcast}]$.
\end{theorem}

\begin{proofT}
  The proof follows from
  Lemmas~\ref{lemma:snapshot-liveness}-\ref{lemma:snapshot-safety}. 
\renewcommand{\toto}{theorem:proof-snapshot}
\end{proofT}

\section{From  SWMR Snapshot to \SCD-broadcast}
\label{sec:from-snapshot-to-SCD}
This section presents an algorithm which builds the \SCD-broadcast
abstraction in  ${\CARW}_{n,t}[\mbox{snapshot]}$. This algorithm
completes the computational equivalence of  snapshot and \SCD-broadcast.
(SWMR snapshot objects can be easily implemented in
${\CAMP}_{n,t}[\mbox{\SCD-broadcast]}$ by instantiating
Algorithm~\ref{algo:snapshot-from-SCD}  with $m=n$, and
only allowing $p_i$  to invoke $\REG.\wwrite(r,-)$.)


\subsection{Algorithm~\ref{algo:Sc-broadcast-from-snapshot}}

\paragraph{Shared objects}
The shared memory is composed of two SWMR snapshot objects (as defined above).
Let $\epsilon$ denote the empty sequence. 
\begin{itemize}
\item
$\SENT[1..n]$: is a snapshot object, initialized to
$[\emptyset, \cdots,\emptyset]$, such that 
$\SENT[i]$ contains the messages scd-broadcast by $p_i$. 
\item
$\SETSEQ[1..n]$: is a snapshot object,  initialized to
  $[\epsilon, \cdots,\epsilon]$, such that $\SETSEQ[i]$ contains
  the sequence of the sets of messages scd-delivered by $p_i$. 
\end{itemize}
The notation $\oplus$ is used for the concatenation of a message set
at the end of a sequence of message sets.

\paragraph{Local objects}
Each process $p_i$manages the following local objects.
\begin{itemize}
\item $\sent_i$ is a  local copy of the snapshot object $\SENT$.
\item  $\setseq_i$ is a  local copy of the snapshot object $\SETSEQ$. 
\item $\todeliveri$ is an auxiliary variable whose aim is to contain the
  next message set that $p_i$ has to scd-deliver. 
\end{itemize}
The function $\members(set\_seq)$
returns the set of all the messages contained in  $set\_seq$.

\paragraph{Description of Algorithm~\ref{algo:Sc-broadcast-from-snapshot}}
%
When a process $p_i$ invokes  $\scbroadcast(m)$, it adds $m$ to
$\sent_i[i]$ and $\SENT[i]$ to inform all the processes 
on the scd-broadcast of $m$. It then invokes the internal procedure
$\progress()$ from which it exits once it has
a set containing $m$ (line~\ref{nRW-01}). 

A background task $T$ ensures that all messages will be
scd-delivered (line~\ref{nRW-02}). This task invokes repeatedly
the internal procedure $\progress()$.
As, locally, both the application process and the underlying task $T$ can
invoke  $\progress()$, which accesses the local variables of $p_i$,
those variables  are protected by a local fair mutual exclusion algorithm
providing the operations $\entermutex()$ and  $\exitmutex()$
(lines~\ref{nRW-03} and~\ref{nRW-11}).

\begin{algorithm}[h!]
\centering{\fbox{
\begin{minipage}[t]{150mm}
\footnotesize 
\renewcommand{\baselinestretch}{2.5}
\resetline
\begin{tabbing}
aaa\=aa\=aaa\=aaaaaa\=\kill

{\bf operation} $\scbroadcast(m)$ {\bf is}\\

\line{nRW-01} \> $\sent_i[i]  \leftarrow \sent_i[i] \cup \{m\}$;
                 $\SENT.\wwrite(\sent_i[i])$;  $\progress()$.~\\~\\


\line{nRW-02} \> {\bf background task} $T$ {\bf is}
                 {\bf repeat forever} $\progress()$  {\bf end repeat}.\\~\\


{\bf procedure} $\progress()$  {\bf is}\\

\line{nRW-03} \>\> $\entermutex()$;\\

\line{nRW-04} \>\>  $\catchup()$;\\

\line{nRW-05} \>\>  $\sent_i \leftarrow \SENT.\snapshot()$;\\

\line{nRW-06} \>\>  $\todeliveri \leftarrow
  (\cup_{1\leq j \leq n}~\sent_i[j]) \setminus \members(\setseq_i[i])$;\\

\line{nRW-07} \>\> {\bf if}  $(\todeliveri \neq\emptyset)$

 {\bf then} \= $\setseq_i[i] \leftarrow \setseq_i[i] \oplus \todeliveri$;\\

\line{nRW-08} \>\>\> 
     $\SETSEQ[i] \leftarrow \setseq_i[i] $;\\
     
\line{nRW-09} \>\>\>   $\scdeliver(\todeliveri)$ \\ 

\line{nRW-10} \>\>  {\bf end if};\\ 

\line{nRW-11} \> \> $\exitmutex()$.\\~\\

 {\bf procedure} $\catchup()$  {\bf is}\\

\line{nRW-12}  \>\> $\setseq_i  \leftarrow \SETSEQ.\snapshot()$;\\

\line{nRW-13}  \>\> {\bf while} \= $(\exists j, set:~ set
\mbox{ is the first set in } \setseq_i[j]:~ set \not\subseteq
\members(\setseq_i[i])$  {\bf do} \\

\line{nRW-14} \>\>\> $\todeliveri \leftarrow set \setminus
                      \members(\setseq_i[i])$;  \\%

\line{nRW-15}\>\>\>
       $\setseq_i[i] \leftarrow \setseq_i[i] \oplus \todeliveri$;
       $\SETSEQ[i] \leftarrow \setseq_i[i]$;\\

\line{nRW-16} \>\>\> $\scdeliver(\todeliveri)$  \\

\line{nRW-17}  \>\> {\bf end while}.

\end{tabbing}
\end{minipage}
}
\caption{An implementation of
         \SCD-broadcast in  $\CARW_{n,t}[\mbox{snapshot}]$ (code for $p_i$)}
\label{algo:Sc-broadcast-from-snapshot}
}
\end{algorithm}

The procedure $\progress()$ first invokes the internal procedure
$\catchup()$, whose aim is to allow $p_i$ to scd-deliver sets of messages
which have been scd-broadcast and not yet locally scd-delivered. 

To this end, $\catchup()$ works as follows (lines~\ref{nRW-12}-\ref{nRW-17}).
Process $p_i$ first obtains a snapshot of $\SETSEQ$, and saves it in $\setseq_i$ 
(line~\ref{nRW-12}). This allows $p_i$ to know which message sets have been
scd-delivered by all the processes; $p_i$ then enters a ``while'' loop to
scd-deliver as many message sets as possible according to what was
scd-delivered by the other processes. 
For each process $p_j$ that has scd-delivered a message set $set$ containing
messages not yet scd-delivered by $p_i$  (predicate of line~\ref{nRW-13}), 
$p_i$ builds a set $\todeliver_i$ containing the messages in $set$ that it
has not  yet scd-delivered  (line~\ref{nRW-14}),  and locally scd-delivers it
(line~\ref{nRW-16}). This local scd-delivery needs to update accordingly
both $\setseq_i[i]$ (local update) and $\SETSEQ[i]$ (global update). 

When it returns from $\catchup()$, $p_i$ strives to scd-deliver messages
not yet scd-delivered by the other processes. To this end, it first obtains
a snapshot of $\SENT$, which it stores in $\sent_i$  (line~\ref{nRW-05}).
If there are messages that can be scd-delivered (computation of $\todeliver_i$
at line~\ref{nRW-06}, and predicate at line~\ref{nRW-07}), $p_i$
scd-delivers them and updates  $\setseq_i[i]$ and  $\SETSEQ[i]$
(lines~\ref{nRW-07}-\ref{nRW-09}) accordingly.

\subsection{Proof of Algorithm~\ref{algo:Sc-broadcast-from-snapshot}}

\begin{lemma}
\label{lemma-broadcast-validity-2}
If a process scd-delivers a set containing a message $m$, some process invoked
$\scbroadcast(m)$. 
\end{lemma}

\begin{proofL}
  The proof follows directly from the text of the algorithm, which copies
  messages from $\SENT$ to $\SETSEQ$, without creating new messages.
\renewcommand{\toto}{lemma-broadcast-validity-2}
\end{proofL}

\begin{lemma}
\label{lemma-broadcast-integrity-2}
No process scd-delivers  the same message twice. 
\end{lemma}

\begin{proofL}
  Let us first observe that, due to lines \ref{nRW-07} and~\ref{nRW-15},
  all messages that are scd-delivered at a process $p_i$ have been  added to
  $\setseq_i[i]$. The proof then follows directly from (a) this observation, 
  (b) the fact that (due to the local mutual exclusion at each process)
  $\setseq_i[i]$ is updated consistently,
  and (c) lines \ref{nRW-06} and~\ref{nRW-14},
  which state that a message already scd-delivered (i.e., a message
  belonging to  $\setseq_i[i]$) cannot be added to $\todeliver_i$. 
\renewcommand{\toto}{lemma-broadcast-integrity-2}
\end{proofL}

\begin{lemma}
\label{lemma-broadcast-termination1-a2}
Any invocation of  $\scbroadcast()$ by a non-faulty process $p_i$ terminates.
\end{lemma}

\begin{proofL}
The proof consists in showing that the internal procedure
$\progress()$ terminates. 
As the  mutex algorithm is assumed to be fair, process $p_i$ cannot block
forever at line~\ref{nRW-03}. Hence, $p_i$ invokes the 
internal  procedure $\catchup()$. It then 
issues first a snapshot invocation on $\SETSEQ$ and stores the 
value it  obtains  the value of $\setseq_i$.
There is consequently a finite number of message sets in $\setseq_i$.
Hence,  the ``while'' of lines~\ref{nRW-13}-\ref{nRW-17}
can be executed only a finite number of times, and it follows that 
any invocation of $\catchup()$ by a non-faulty process terminates. 
The same reasoning (replacing $\SETSEQ$ by $\SENT$) shows that
process $p_i$ cannot block forever when it executes the 
lines~\ref{nRW-05}-\ref{nRW-10} of the procedure $\progress()$. 
\renewcommand{\toto}{lemma-broadcast-termination1-a2}
\end{proofL}

\begin{lemma}
\label{lemma-broadcast-termination1-b2}
If a non-faulty process scd-broadcasts a message $m$,
it scd-delivers a message set containing $m$.
\end{lemma}

\begin{proofL}
Let $p_i$ be a non-faulty process that  scd-broadcasts a message $m$.
As it is non-faulty, $p_i$ adds $m$ to  $\SENT[i]$
and then invokes  $\progress()$  (line~\ref{nRW-01}).
As $m\in \SENT$, it is eventually added to $\todeliveri$
if not yet scd-delivered (line~\ref{nRW-06}),
and scd-delivered at line~\ref{nRW-09}, which concludes the proof
of the lemma. 
\renewcommand{\toto}{lemma-broadcast-termination1-b2}
\end{proofL}

\begin{lemma}
\label{lemma-broadcast-termination2-2}
If a non-faulty process scd-delivers a message $m$, every  non-faulty process
scd-delivers a message set containing $m$.
\end{lemma}

\begin{proofL}
Let us assume that a process scd-delivers a message set containing
a message $m$. It follows that the  process that invoked $\scbroadcast(m)$ 
added $m$ to $\SENT$ (otherwise no process could scd-deliver $m$). 
Let $p_i$ be a correct process. 
It invokes $\progress()$ infinitely often  (line~\ref{nRW-02}).
Hence, there is a first execution of $\progress()$
such that $sent_i$ contains $m$   (line~\ref{nRW-05}).
If then follows from line~\ref{nRW-06}  that $m$ will be added to
$\todeliver_i$ (if not yet scd-delivered). If follows that $p_i$ will
scd-deliver a set of messages containing $m$ at line~\ref{nRW-09}.    
\renewcommand{\toto}{lemma-broadcast-termination2-2}
\end{proofL}


\begin{lemma}
\label{lemma-broadcast-sc-ordering-2}
Let $p_i$ be a process that scd-delivers a set $ms_i$ containing a
message $m$ and later scd-delivers a set $ms'_i$ containing a message
$m'$.  No process $p_j$ scd-delivers first a set $ms'_j$ containing $m'$
and later a set $ms_j$ containing $m$.
\end{lemma}
\begin{proofL}
  Let us consider two messages $m$ and $m'$. Due to
   total order property on the  operations on  the  snapshot object
  $\SENT$, it is possible to order the write operations
  of $m$ and $m'$ into $\SENT$. Without loss of generality, let us assume
  that $m$ is added to $\SENT$ before $m'$.
  We show that no process scd-delivers $m'$ before $m$.\footnote{Let us
    notice that it is possible that a process scd-delivers them in two
    different message sets, while another process scd-delivers them in the same set
    (which does not contradicts the lemma).}

Let us consider a process $p_i$ that scd-delivers the message $m'$.
There are two cases. 
\begin{itemize}
\item $p_i$ scd-delivers the message $m'$  at line~\ref{nRW-09}. 
Hence, $p_i$ obtained $m'$ from the snapshot object $\SENT$
(lines~\ref{nRW-05}-\ref{nRW-06}).
As $m$ was written in $\SENT$ before $m'$, we conclude that
$\SENT$ contains $m$. It then follows from line~\ref{nRW-06} that,
if $p_i$ has not scd-delivered $m$ before (i.e., $m$ is not in  
$\setseq_i[i]$), then $p_i$ scd-delivers it in the same set as $m'$. 
\item $p_i$ scd-delivers the message $m'$  at line~\ref{nRW-16}.
  Due to the predicate used at line~\ref{nRW-13} to build a set
  of message to scd-deliver, this means that there is a process $p_j$
  that has previously scd-delivered a set of messages containing $m'$.\\
  Moreover, let us observe that the first time the  message $m'$ is copied
  from $\SENT$ to some $\SETSEQ [x]$ occurs at line~\ref{nRW-08}.
  As $m$ was written in $\SENT$ before $m'$, the corresponding process
  $p_x$ cannot see $m'$ and not $m$. 
  It follows from the previous item that $p_x$ has scd-delivered $m$
  in the same  message set (as the one including $m'$), or in a previous 
  message set. It then follows from the predicate of line~\ref{nRW-13} that
  $p_i$ cannot scd-delivers $m'$ before $m$.

  To summarize, the scd-deliveries of message sets in the procedure
  $\catchup()$  cannot violate the MS-Ordering property, which is established
  at lines~\ref{nRW-06}-\ref{nRW-10}. 
\end{itemize}
\renewcommand{\toto}{lemma-broadcast-sc-ordering-2}
\end{proofL}

\begin{theorem}
\label{theorem:sc-broadcast-from-RW}
Algorithm~{\em{\ref{algo:Sc-broadcast-from-snapshot}}}
implements the {\em \SCD-Broadcast} abstraction in the system model
$\CARW_{n,t}[t<n]$.
\end{theorem}
\begin{proofT}
  The proof follows from
  Lemma~\ref{lemma-broadcast-validity-2} (Validity), 
  Lemma~\ref{lemma-broadcast-integrity-2} (Integrity),
  Lemmas~\ref{lemma-broadcast-termination1-a2}
  and~\ref{lemma-broadcast-termination1-b2} (Termination-1), 
  Lemma~\ref{lemma-broadcast-termination2-2} (Termination-2), and
  Lemma~\ref{lemma-broadcast-sc-ordering-2} (MS-Ordering).
\renewcommand{\toto}{theorem:sc-broadcast-from-RW}
\end{proofT}


\section{Conclusion}
\label{sec:conclusion}
This paper has introduced a new communication abstraction
(\SCD-broadcast) providing processes
with an abstraction level between reliable broadcast and total
order broadcast (which captures
the necessary and sufficient constraint on message
deliveries which allows  consensus objects to  be implemented in
asynchronous crash-prone message-passing systems).

More precisely, \SCD-broadcast  captures
the  abstraction level which is  ``necessary and sufficient''
to implement read/write registers and snapshot objects on top of
asynchronous message-passing systems prone to process
failures. ``Sufficient'' means here that no other notion or
object\footnote{The notion of intersecting quorums
is neither  provided by the abstraction level offered by \SCD-broadcast,
nor required --in addition to \SCD-broadcast--
to implement registers or snapshot objects. Actually, 
it is  hidden  and majority quorums appear only in the implementation
of \SCD-broadcast.}
is needed to build a register or a snapshot object 
at the abstraction level provided by
\SCD-broadcast, while ``necessary'' means that the objects that are
built (registers and snapshot objects) are the weakest from a shared
memory computational point of view.  


As announced in the Introduction, an algorithm implementing
\SCD-broadcast in an asynchronous message-passing system where any minority
of processes may crash is described in Appendix~\ref{sec:algo-for-SC}.
This algorithm  requires $O(n^2)$ protocol messages per invocation of
$\scbroadcast()$.  It follows that the \SCD-broadcast-based
MWMR snapshot algorithm
presented in the paper requires $O(n^2)$ protocol messages
per invocation of $\snapshot()$ or $\wwrite()$ operation.
This is the best read/write snapshot algorithm we know in the context
of asynchronous message-passing systems.

\newpage
\setcounter{page}{1}
\pagenumbering{roman}

\section*{Acknowledgments}
This work has been partially supported by the  Franco-German
DFG-ANR Project 40300781 DISCMAT (devoted to connections between
mathematics and distributed computing), and the French ANR project DESCARTES 
(devoted to layered and modular structures in distributed computing). 
The authors want to thank Faith Ellen for fruitful exchanges on
shared memory snapshot.


\newpage
\appendix

\section{An Implementation of \SCD-broadcast  in Message-Passing Systems}
\label{sec:algo-for-SC}
This section  shows that the \SCD-broadcast communication abstraction is not an
oracle-like object which allows us to extend our understanding of computing,
but cannot be implemented. It describes an  implementation of  \SCD-broadcast 
in $\CAMP_{n,t}[t<n/2]$, which is the weakest assumption on process failures
that allows a  read/write register to be built on top of an asynchronous 
message-passing system~\cite{ABD95} (see footnote~\ref{footnote1}).

To simplify the presentation, and without loss of generality, we consider
that the communication channels are FIFO. The associated communication
operations are denoted $\fifobroadcast()$ and $\fifodeliver()$. 

\subsection{Algorithm~\ref{algo:Sc-broadcast-in-RW}}

\paragraph{Local variables at a process $p_i$}
Each process $p_i$ manages the following local
variables. 
\begin{itemize}     
\item $\bufferi$: buffer where are stored the messages not yet
scd-delivered in a message set. 
\item $\todeliveri$: next set of messages to be scd-delivered.
\item $sn_i$: local sequence number (initialized to $0$),
              which measures the local progress of $p_i$.
\item $clock_i[1..n]$: array of sequence numbers.
  $clock_i[j]$ is the greatest sequence number $x$ such that the  application
  message  identified by $\langle x,j\rangle$ was in a message set scd-delivered
  by $p_i$.  
\end{itemize}

\paragraph{Operation $\scbroadcast()$}
When $p_i$ invokes $\scbroadcast(m)$, where $m$ is an application message,
it sends the message \FORWARD$(m,i,sn_i,i,sn_i)$ to itself
(this simplifies the writing of the algorithm), and waits until
it has no more message from itself pending in $\buffer_i$, which means
it has scd-delivered a set containing $m$.

A protocol message \FORWARD$()$ (line~\ref{SCD-from-MP-01}) is made up of five
fields: the associated application message $m$, and two pairs, each
made up of a sequence number and a process identity.  The first pair
($sd,sn)$ is the identity of the application message, while the second
one $(f,\snf)$ is the local progress ($\snf$) of the forwarder process
${\mathit{p_f}}$ when it forwards this protocol message.

\paragraph{Reception of  \FORWARD$(m,sd,\snsd,f,\snf)$}
When a process $p_i$ receives such a protocol message, it first
invokes $\forward(m,sd,\snsd,f,\snf)$ to participate in the reliable
broadcast of this message (line~\ref{SCD-from-MP-03}), and then
invokes $\trydeliver()$ to see if a message set can be scd-delivered
(line~\ref{SCD-from-MP-04}).

\paragraph{Procedure $\forward()$}
This procedure can be seen as an enrichment (with the fields $f$ and $\snf$)
of the reliable broadcast implemented by the messages
\FORWARD$(m,sd,\snsd,-,-)$.
Considering such a message \FORWARD$(m,sd,\snsd,f,\snf)$, 
$m$ was scd-broadcast by $p_{sd}$ at its local time $\snsd$, and
relayed by the forwarding process  $p_f$ at its local time $\snf$.
If $\snsd \leq  clock_i[sd]$, $p_i$ has already  scd-delivered a message set
containing $m$ (see lines~\ref{SCD-from-MP-18} and \ref{SCD-from-MP-20}). 
If $\snsd > clock_i[sd]$, there are two cases. 
\begin{itemize}
\item
  The message $m$ is not in $\bufferi$.
  In this case, $p_i$ creates a quadruplet $msg$, and  adds it to 
  $\bufferi$ (lines~\ref{SCD-from-MP-08}-\ref{SCD-from-MP-10}). This quadruplet 
  $\langle msg.m, msg.sd,msg.f,msg.cl\rangle$ is such that
\begin{itemize}
\item 
the field $msg.m$ contains the  application message $m$,
\item 
the field $msg.sd$ contains the id of the sender of this application message,
\item
the field  $msg.sn$ contains the local date associated with $m$ by its sender, 
\item
the field $msg.cl$ is an array of size $n$, such that $msg.cl[x]$ = sequence
number (initially $+\infty$)
associated with $m$ by $p_x$ when it broadcast \FORWARD$(msg.m,-,-,-,-)$.
This last field is crucial in the scd-delivery of a message set containing $m$.
\end{itemize}
After the quadruplet $msg$ has been built, $p_i$ first adds it to
$\bufferi$ (line~\ref{SCD-from-MP-10}), and invokes
(line~\ref{SCD-from-MP-11})
$\fifobroadcast$~\FORWARD$(m,sd,\snsd,i,sn_i)$ to implement the
reliable broadcast of $m$ identified by $\langle sd,\snsd\rangle$.
Finally, $p_i$ records its progress by increasing $sn_i$
(line~\ref{SCD-from-MP-12}).
\item
There is a quadruplet  $msg$ in $\bufferi$ 
associated with $m$, i.e., $msg=\langle m, sd,-, -\rangle \in \bufferi$
(predicate of line~\ref{SCD-from-MP-06}). In this case,  $p_i$  assigns $\snf$
to $msg.cl[f]$ (line~\ref{SCD-from-MP-07}), thereby indicating that
$m$ was known and forwarded by $\mathit{p_f}$ at its local time $\snf$.
\end{itemize}

\begin{algorithm}[h!]
\centering{\fbox{
\begin{minipage}[t]{150mm}
\footnotesize 
\renewcommand{\baselinestretch}{2.5}
\resetline
\begin{tabbing}
aaa\=aa\=aaa\=aaaaaa\=\kill

{\bf operation} $\scbroadcast(m)$ {\bf is}\\

\line{SCD-from-MP-01}  \>\> $\forward(m,i,sn_i,i,sn_i)$; \\
\line{SCD-from-MP-02}  \>\> $\wait (\nexists ~msg\in \bufferi : msg.sd = i)$.\\
~\\
~

{\bf when the message} \= \FORWARD$(m,sd,\snsd,f,\snf)$
          {\bf is fifo-delivered} {\bf do}   $~~$ \% from $\mathit{p_f}$ \\

\line{SCD-from-MP-03}  \>\> $\forward(m,sd,\snsd,f,\snf)$; \\
\line{SCD-from-MP-04}  \>\> $\trydeliver()$.\\~\\


{\bf procedure} $\forward(m,sd,\snsd,f,\snf)$  {\bf is}\\

\line{SCD-from-MP-05} \>\> {\bf if} \=   $(\snsd>clock_i[sd])$\\ 

\line{SCD-from-MP-06} \>\>\> {\bf then} \= {\bf if} \=
$(\exists~ msg\in \bufferi: msg.sd = sd \land msg.sn = \snsd)$\\

\line{SCD-from-MP-07} \>\>\>\>\> {\bf then} \=  $msg.cl[f] \leftarrow \snf$\\

\line{SCD-from-MP-08} \>\>\>\>\> {\bf else} \>
          $threshold[1..n]  \leftarrow [\infty,\dots,\infty]$;
          $threshold[f]  \leftarrow \snf$;\\

\line{SCD-from-MP-09} \>\>\>\>\>\>
     {\bf let} $msg \leftarrow\langle m, sd, \snsd, threshold[1..n]\rangle$;\\
     
\line{SCD-from-MP-10} \>\>\>\>\>\>$\bufferi \leftarrow \bufferi\cup \{msg\}$;\\ 
\line{SCD-from-MP-11} \>\>\>\>\>\>
                             $\fifobroadcast$~\FORWARD$(m,sd,\snsd,i,sn_i)$;\\ 
\line{SCD-from-MP-12} \>\>\>\>\>\> $sn_i \leftarrow sn_i + 1$\\ 
\line{SCD-from-MP-13} \>\>\>\>  {\bf end if}\\ 
\line{SCD-from-MP-14} \>\>  {\bf end if};\\~\\ 

{\bf procedure} $\trydeliver()$  {\bf is}\\

\line{SCD-from-MP-15} \>\> \textbf{let} \= $\todeliveri \leftarrow
\{msg \in \bufferi: |\{f : msg.cl[f] < \infty\}| > \frac{n}{2}\} $;\\

\line{SCD-from-MP-16} 
\textbf{while} \=
     $(\exists msg\in \todeliveri, msg'\in \bufferi\setminus
     \todeliveri : |\{f : msg.cl[f] < msg'.cl[f] \}| \le \frac{n}{2}$)
     \textbf{do}\\
  
\>\>\> $\todeliveri\leftarrow \todeliveri\setminus \{msg\}$\\
   
\>\> \textbf{end while};\\

\line{SCD-from-MP-17} \>\> {\bf if} \= $(\todeliveri \neq \emptyset)$\\

\line{SCD-from-MP-18} \>\>\> {\bf then} \=
  {\bf for each} $(msg\in \todeliveri$ such that $clock_i[msg.sd] < msg.sn)$\\
 \>\>\>\> $~~~~~~~~~~~~~~~~$
    {\bf do} $clock_i[msg.sd] \leftarrow msg.sn$ {\bf end for}; \\
     
    \line{SCD-from-MP-19} \>\>\>\>
                         $\bufferi\leftarrow \bufferi\setminus \todeliveri$;\\
\line{SCD-from-MP-20} \>\>\>\>
     $ms \leftarrow\{m : \exists~ msg\in \todeliveri : msg.m = m\}$;
    $\scdeliver(ms)$\\

\line{SCD-from-MP-21} \>\> {\bf end if}.

\end{tabbing}
\end{minipage}
}
\caption{An implementation of
         \SCD-broadcast in  $\CAMP_{n,t}[t<n/2]$ (code for $p_i$)}
\label{algo:Sc-broadcast-in-RW}
}
\end{algorithm}

\paragraph{Procedure $\trydeliver()$}
When it executes $\trydeliver()$, $p_i$ first computes  the set
$\todeliveri$ of the quadruplets $msg$ containing  application
messages $m$ which have been seen by a majority of processes
(line~\ref{SCD-from-MP-15}).  From $p_i$'s point of view, a message has been
seen by a process $\mathit{p_f}$ if $msg.cl[f]$ has been set to a
finite value (line~\ref{SCD-from-MP-07}).

If a majority of processes received first a message
\FORWARD$(m',-,-,-,-)$ and later another message \FORWARD$(m,-,-,-,-)$,
it might be
that some process $p_j$ scd-delivered a set containing $m'$ before
scd-delivering a set containing $m$.  Therefore, $p_i$ must avoid 
scd-delivering a set containing $m$ before scd-delivering a set containing $m'$.
This is done at line~\ref{SCD-from-MP-16}, where $p_i$ withdraws the
quadruplet $msg$ corresponding to $m$ if it has not enough information
to deliver $m'$ (i.e. the corresponding $msg'$ is not in
$\todeliver_i$) or it does not have the proof that the situation cannot 
happen, i.e. no majority of processes saw the message
corresponding to $msg$ before the message corresponding to $msg'$.

If $\todeliveri$ is not empty after it has been purged
(lines~\ref{SCD-from-MP-16}-\ref{SCD-from-MP-17}), $p_i$ computes a
message set to scd-deliver.  This set $ms$ contains all the application
messages in the quadruplets of $\todeliveri$
(line~\ref{SCD-from-MP-20}).  These quadruplets are withdrawn from
$\bufferi$ (line~\ref{SCD-from-MP-18}).  Moreover, before this
scd-delivery, $p_i$ needs to updates $clock_i[x]$ for all the entries
such that $x=msg.sd$ where $msg\in \todeliveri$
(line~\ref{SCD-from-MP-18}).  This update is needed to ensure that
the future uses of the predicate of line~\ref{SCD-from-MP-17} are correct.

\subsection{Proof of Algorithm~\ref{algo:Sc-broadcast-in-RW}}

\begin{lemma}
\label{lemma-broadcast-validity}
If a process  scd-delivers a set containing  $m$, some
process invoked $\scbroadcast(m)$.
\end{lemma}

\begin{proofL}
  If process $p_i$ scd-delivers a set containing a message $m$, it
  has previously added into $\bufferi$ a quadruplet $msg$
  such that $msg.m=m$ (line \ref{SCD-from-MP-10}), for which it
  has fifo-received at least $\frac{n}{2}$ \FORWARD$(m,-,-,-,-)$
  messages. The first of these messages ever sent was sent
  after a process invoked $\scbroadcast(m)$.
  \renewcommand{\toto}{lemma-broadcast-validity}
\end{proofL}

\begin{lemma}
\label{lemma-broadcast-integrity}
No process scd-delivers  the same message twice. 
\end{lemma}

\begin{proofL}
  After a message $m$ scd-broadcast by $p_{sd}$ with a sequence number
  $\snsd$ is scd-delivered by $p_i$, $clock_i[sd] \ge \snsd$ thanks
  to line \ref{SCD-from-MP-18} and there is no $msg\in \buffer_i$ with
  $msg.sd=sd$ and $msg.sn=\snsd$, as it was removed on line
  \ref{SCD-from-MP-19}.  Thanks to line \ref{SCD-from-MP-05}, no such
  $msg'$ will be added again in $\bufferi$. As $\todeliveri$ is
  defined as a subset of $\bufferi$ on line \ref{SCD-from-MP-15}, $m$
  will never be scd-delivered by $p_i$ again.
  \renewcommand{\toto}{lemma-broadcast-integrity}
\end{proofL}

\begin{lemma} \label{lemma:broadcast}
If a message \FORWARD$(m, sd, \snsd,i, sn_i)$ is broadcast by a non-faulty
process $p_i$, then each non-faulty process $p_j$ broadcasts a single
message \FORWARD$(m, sd, \snsd,j, sn_j)$.
\end{lemma}
\begin{proofL}
  First, we prove that $p_j$ broadcasts a message
  \FORWARD$(m, sd,\snsd,j, sn_j)$.
  As $p_i$ is non-faulty, $p_j$ will eventually receive the
message sent by $p_i$. At that time, if $\snsd > clock_j[sd]$, after
the condition on line~\ref{SCD-from-MP-06} and whatever its result,
$\buffer_i$ contains a value $msg$ with $msg.sd = sd$ and $msg.\snsd =
\snsd$. That $msg$ was inserted at line~\ref{SCD-from-MP-10} (possibly
after the reception of a different message), just before $p_j$ sent a
message \FORWARD$(m, sd, \snsd, j,sn_j)$ at line~\ref{SCD-from-MP-11}.
Otherwise, $clock_j[sd]$ was incremented on line~\ref{SCD-from-MP-18},
when validating some $msg'$  added to $\bufferj$ after $p_j$
received a (first) message \FORWARD$(msg'.m,sd,\snsd,f, clock_f[sd])$
from $p_f$. Because the messages \FORWARD$()$ are fifo-broadcast
(hence they are  delivered in their sending order), $p_{sd}$ sent
message \FORWARD$(msg.m, sd, \snsd,sd, \snsd)$ before
\FORWARD$(msg'.m, sd, clock_j[sd],  sd,clock_j[sd])$,
and all other processes only forward
messages, $p_j$ received a message \FORWARD$(msg.m, sd, \snsd,-,-)$
from $p_f$ before the message \FORWARD$(msg'.m, sd, clock_j[sd],-,-)$.
At that time, $\snsd > clock_j[sd]$, so the previous case applies.

After $p_j$ broadcasts its message \FORWARD$(m, sd, \snsd,j, sn_j)$ on
line~\ref{SCD-from-MP-11}, there is a $msg\in \buffer_j$ with $ts(msg)
= \langle sd, \snsd \rangle$, until it is removed on
line~\ref{SCD-from-MP-16} and $clock_j[sd] \ge \snsd$. Therefore, one of the
conditions at lines~\ref{SCD-from-MP-05} and~\ref{SCD-from-MP-06} will
stay false for the stamp $ts(msg)$ and $p_j$ will never execute
line~\ref{SCD-from-MP-11} with the same stamp $\langle sd, \snsd\rangle$ later.
\renewcommand{\toto}{lemma:broadcast}
\end{proofL}

\begin{lemma}
  \label{lemma-broadcast-sc-ordering}
  Let $p_i$ be a process that  scd-delivers a set $ms_i$ containing a
  message $m$ and later  scd-delivers a set $ms'_i$ containing a message
  $m'$.  No process $p_j$  scd-delivers first a set $ms'_j$ containing
  $m'$  and later a set $ms_j$ containing $m$.
\end{lemma}

\begin{proofL}
  Let us suppose there are two messages $m$ and $m'$ and two processes
  $p_i$ and $p_j$ such that $p_i$ scd-delivers a set $ms_i$
  containing $m$ and later scd-delivers a set $ms'_i$ containing $m'$
  and $p_j$ scd-delivers a set $ms'_j$ containing $m'$ and later
  scd-delivers a set $ms_j$ containing $m$.

  When $m$ is delivered by $p_i$, there is an element $msg\in
  \buffer_i$ such that $msg.m = m$ and because of line
  \ref{SCD-from-MP-15}, $p_i$ has received a message \FORWARD$(m,-,-,-,-)$
  from more than $\frac{n}{2}$ processes.
  \begin{itemize}
  \item If there is no element $msg'\in \buffer_i$ such that $msg'.m =
    m'$, since $m'$ has not been delivered by $p_i$ yet, $p_i$ has not
    received a message \FORWARD$(m',-,-,-,-)$ from any process
    (lines \ref{SCD-from-MP-10} and \ref{SCD-from-MP-19}). Therefore,
    because the communication channels are FIFO, more than
    $\frac{n}{2}$ processes have sent a message \FORWARD$(m,-,-,-,-)$
    before sending a message \FORWARD$(m',-,-,-,-)$.
   \item Otherwise, $msg'\notin \todeliver_i$ after
    line~\ref{SCD-from-MP-16}. As the communication channels are
    FIFO, more than $\frac{n}{2}$ processes have sent a message
    \FORWARD$(m,-,-,-,-)$ before a message \FORWARD$(m',-,-,-,-)$.
  \end{itemize}

  Using the same reasoning, it follows that
  when $m'$ is delivered by $p_j$, more than
  $\frac{n}{2}$ processes have sent a message \FORWARD$(m',-,-,-,-)$ before
  sending a message \FORWARD$(m,-,-,-,-)$.  There exists a process $p_k$ in
  the intersection of the two majorities, that has both sent a message
  \FORWARD$(m',-,-,-,-)$ before sending \FORWARD$(m,-,-,-,-)$ and sent a message
  \FORWARD$(m',-,-,-,-)$ before sending \FORWARD$(m,-,-,-,-)$.
  However, by Lemma~\ref{lemma:broadcast}, $p_k$ can only send one message
  \FORWARD$(m',-,-,-,-)$ and one message \FORWARD$(m,-,-,-,-)$, which leads
  to a  contradiction.
\renewcommand{\toto}{lemma-broadcast-sc-ordering}
\end{proofL}

\begin{figure}[h!]
  \begin{center}
    \begin{tikzpicture}

      \draw[->] (0,0) node[left]{$p_i$} -- (11,0);
      \draw[->] (0,1.5) node[left]{$p_f$} -- (11,1.5);

      \draw (2.75,2.1) node{\footnotesize$\scbroadcast(m_k)$};
      \draw (3.4,0.75) node[left]{\footnotesize\FORWARD$(m_k,f,sn_f(k),f,sn_f(k))$};
      \draw (10.2,0.75) node[right]{$\cdots$};

      \draw[fill=white] (2,1.2) rectangle (3.5,1.8);
      \draw (2.1,1.5) node{$\bullet$};
      \draw[-latex] (2.3,1.5) to[out=0,in=120,distance=20] (3.8,0.1);
      \draw[-latex] (2.2,1.6) to[out=60,in=120,distance=5] (2.5,1.6);

      \draw[fill=white] (4.8,1.2) rectangle (7.5,1.8);
      \draw (4.9,1.5) node{$\bullet$};
      \draw[-latex] (5,1.4) -- (6,0.1);
      \draw[-latex] (5,1.6) to[out=60,in=120,distance=5] (5.3,1.6);

      \draw[fill=white] (7.8,1.2) rectangle (9.5,1.8);
      \draw (7.9,1.5) node{$\bullet$};
      \draw[-latex] (8,1.4) -- (9,0.1);
      \draw[-latex] (8,1.6) to[out=60,in=120,distance=5] (8.3,1.6);

      \draw (5.5,0.75) node[right]{\footnotesize$sn_f(k1)$};
      \draw (8.5,0.75) node[right]{\footnotesize$sn_f(k2)$};

      \draw[->] (4.8,-0.3) -- (5.8,-0.3);
      \draw (5.9,-0.3) node{${}_i^\star$};
      \draw[->] (6.1,-0.3) -- (8.8,-0.3);
      \draw (8.9,-0.3) node{${}_i^\star$};
      \draw[->] (8.8,-0.6) -- (4.8,-0.6);
      \draw (4.7,-0.6) node{${}_i^\star$};

      \draw (1.5,-0.9) node[below]{\footnotesize\FORWARD$(m,sd,\snsd,-,-)$};
      \draw (4,-1.3) node[below]{\footnotesize\FORWARD$(m,sd,\snsd,-,-)$};
      \draw[-latex,dashed,thick] (1.5,-0.9) -- (1.8,-0.1);
      \draw[-latex,dashed,thick] (4,-1.3) -- (4.2,1.4);
      \draw[-latex,dashed,thick] (4.3,1.4) -- (4.7,0.1);
      \draw[-latex,dashed,thick] (4.3,1.6) to[out=60,in=120,distance=5] (4.6,1.6);

      \draw (7.1,2.5) node[above]{\footnotesize\FORWARD$(m_{l+1},sd_{l+1},sn_{sd_{l+1}},-,-)$};
      \draw[-latex,dotted,thick] (7.0,2.5) -- (7.2,1.55);
      \draw[-latex,dotted,thick] (7.1,2.5) -- (7.3,1.55);
      \draw[-latex,dotted,thick] (7.2,2.5) -- (7.4,1.55);
      \draw[-latex,dotted,thick] (7.2,1.45) -- (7.4,0.05);
      \draw[-latex,dotted,thick] (7.3,1.45) -- (7.5,0.05);
      \draw[-latex,dotted,thick] (7.4,1.45) -- (7.6,0.05);

    \end{tikzpicture}
  \end{center}
  \caption{Message pattern introduced in Lemma~\ref{lemma:liveness}}
  \label{fig:lemma:liveness}
\end{figure}
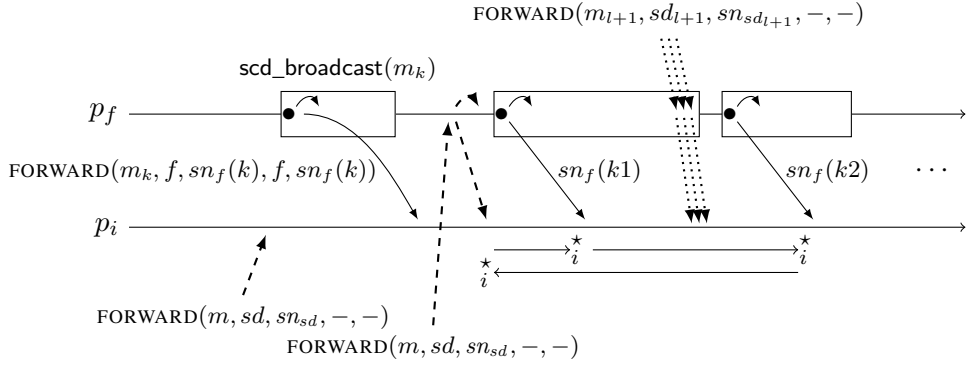

\begin{lemma} \label{lemma:liveness}
  If a message \FORWARD$(m, sd, \snsd,i, sn_i)$ is fifo-broadcast
  by a non-faulty
  process $p_i$, this process  scd-delivers a set containing $m$.
\end{lemma}

\begin{proofL}
  Let $p_i$ be a non-faulty process.   For any pair of messages
  $msg$ and $msg'$ ever inserted
  in $\bufferi$,  let $ts = ts(msg)$ and  $ts' = ts(msg')$.
  Let  $\rightarrow_i$ be the dependency relation defined  as follows:
  $ts \rightarrow_i ts' \stackrel{def}{=}|\{j : msg'.cl[j] < msg.cl[j]
  \}| \le \frac{n}{2}$ (i.e. the dependency does not exist if $p_i$
  knows that a majority of processes have seen the first update
  --due to $msg'$-- before the second --due to $msg$--).
  Let $\rightarrow_i^\star$ denote the transitive  closure of $\rightarrow_i$.
  
  Let us suppose (by contradiction) that the timestamp $\langle sd,
  \snsd\rangle$ associated with the message $m$ (carried by the
  protocol message \FORWARD$(m, sd, \snsd,i, sn_i)$ fifo-broadcast by
  $p_i$), has an infinity of predecessors according to
  $\rightarrow_i^\star$.  As the number of processes is finite, an
  infinity of these predecessors have been generated by the same
  process, let us say $p_f$.  Let $\langle f, sn_f(k) \rangle_{k\in
    \mathbb{N}}$ be the infinite sequence of the timestamps associated
  with the invocations of the $\scbroadcast()$ issued by $p_f$. The
  situation is depicted by Figure~\ref{fig:lemma:liveness}.
  
  As $p_i$ is non-faulty, $p_f$ eventually receives a message
  \FORWARD$(m,sd, \snsd, i, sn_i)$, which means $p_f$ broadcast an
  infinity of messages \FORWARD$(m(k), f, sn_f(k), f, sn_f(k))$ after
  \FORWARD$(m, sd, \snsd, f, sn_f)$. Let $\langle f, sn_f(k1) \rangle$
  and $\langle f, sn_f(k2) \rangle$ be the timestamps associated with the
  next two messages sent by $p_f$, with $sn_f(k1) < sn_f(k2)$.  By
  hypothesis, we have $\langle f, sn_f(k2) \rangle\rightarrow_i^\star
  \langle sd, \snsd \rangle$.  Moreover, all processes received their
  first message \FORWARD$(m, sd, \snsd, -,-)$ before their first
  message \FORWARD$(m(k), f, sn_f(k), -,-)$, so $\langle sd, \snsd
  \rangle\rightarrow_i^\star \langle f, sn_f(k1) \rangle$. Let us
  express the path $\langle f, sn_f(k2) \rangle\rightarrow_i^\star
  \langle f, sn_f(k1) \rangle$:\\ $\langle f, sn_f(k2) \rangle =
  \langle sd'(1), sn'(1) \rangle \rightarrow_i \langle sd'(2), sn'(2)
  \rangle \rightarrow_i \dots \rightarrow_i \langle sd(m), sn'(m)
  \rangle = \langle f, sn_f(k1) \rangle$.

  In the time interval starting when $p_f$ sent the message
  \FORWARD$(m(k1), f, sn_f(k1), f, sn_f(k1))$ and finishing when it
  sent the the message \FORWARD$(m(k2), f, sn_f(k2), f, sn_f(k2))$,
  the waiting condition of line~\ref{SCD-from-MP-02} became true, so $p_f$
  scd-delivered a set containing the message $m(k1)$, and according
  to Lemma~\ref{lemma-broadcast-validity}, no set containing the
  message $m(k2)$. Therefore, there is an index $l$ such that process
  $p_f$ delivered sets containing messages associated with a timestamp
  $\langle sd'(l), sn'(l) \rangle$ for all $l'>l$ but not for
  $l'=l$. Because the channels are FIFO and thanks to
  lines~\ref{SCD-from-MP-15} and~\ref{SCD-from-MP-16}, it means that a
  majority of processes have sent a message \FORWARD$(-, sd'(l+1),
  sn'(l+1),-,-)$ before a message \FORWARD$(-, sd'(l), sn'(l),-,-)$,
  which contradicts the fact that $\langle sd'(l), sn'(l) \rangle
  \rightarrow_i \langle sd'(l+1), sn'(l+1) \rangle$.

  Let us suppose a non-faulty process $p_i$ has fifo-broadcast a
  message \FORWARD$(m, sd, \snsd,i, sn_i)$
  (line~\ref{SCD-from-MP-10}).  It inserted a quadruplet $msg$ with
  timestamp $\langle sd, \snsd\rangle$ on line~\ref{SCD-from-MP-09}
  and by what precedes, $\langle sd, \snsd\rangle$ has a finite number
  of predecessors $\langle sd_1, sn_1\rangle, \dots, \langle sd_l,
  sn_l\rangle $ according to $\rightarrow_i^\star$.  As $p_i$ is
  non-faulty, according to Lemma~\ref{lemma:broadcast}, it eventually
  receives a message \FORWARD$(-,sd_k, sn_k,-,-)$ for all $1\le k \le
  l$ and from all non-faulty processes, which are in majority.

  Let $\mathit{pred}$ be the set of all quadruplets $msg'$ such that
  $\langle msg'.sd, msg'.\snsd \rangle \rightarrow_i^\star \langle sd,
  \snsd \rangle$.  Let us consider the moment when $p_i$ receives the
  last message \FORWARD$(-,sd_k, sn_k, f, sn_f)$ sent by a correct
  process $p_f$.  For all $msg'\in \mathit{pred}$, either $msg'.m$ has
  already been delivered or $msg'$ is inserted $\todeliver_i$ on
  line~\ref{SCD-from-MP-15}.  Moreover, no $msg'\in \mathit{pred}$
  will be removed from $\todeliver_i$, on line~\ref{SCD-from-MP-16},
  as the removal condition is the same as the definition of
  $\rightarrow_i$.  In particular for $msg' = msg$, either $m$ has
  already been scd-delivered or $m$ is present in $\todeliver_i$ on
  line~\ref{SCD-from-MP-17} and will be scd-delivered on
  line~\ref{SCD-from-MP-20}.  \renewcommand{\toto}{lemma:liveness}
\end{proofL}

\begin{lemma}
\label{lemma-broadcast-termination1}
If a non-faulty process scd-broadcasts a message $m$, it scd-delivers
a message set containing $m$.
\end{lemma}

\begin{proofL}
  If a non-faulty process scd-broadcasts a message $m$, it sends a
  message \FORWARD$(m, i, \snsd,i,\snsd)$ on
  line~\ref{SCD-from-MP-11}, so it scd-delivers a message set
  containing $m$ by lemma~\ref{lemma:liveness}.
  \renewcommand{\toto}{lemma-broadcast-termination1}
\end{proofL}

\begin{lemma}
\label{lemma-broadcast-termination2}
If a non-faulty process scd-delivers a message $m$, every  non-faulty process
scd-delivers a message set containing $m$.
\end{lemma}

\begin{proofL}
  Suppose a non-faulty process $p_i$ scd-delivers a message $m$. At
  line~\ref{SCD-from-MP-20},
  there is $msg\in \todeliver_i$ such that $msg.m = m$.
  At line~\ref{SCD-from-MP-15}, $msg\in \buffer_i$, and 
  $msg$ was inserted in $\buffer_i$ at line~\ref{SCD-from-MP-10},
  just before $p_i$ sent message \FORWARD$(m, sd, \snsd,i, sn_i)$.
  By Lemma~\ref{lemma:broadcast}, every non-faulty process $p_j$ sends
  a message \FORWARD$(m, sd, \snsd,j, sn_j)$, so by Lemma~\ref{lemma:liveness},
  $p_j$ scd-delivers a message set containing $m$.
  \renewcommand{\toto}{lemma-broadcast-termination2}
\end{proofL}

\begin{theorem}
\label{theorem:SCD}
Algorithm~{\em\ref{algo:Sc-broadcast-in-RW}}
implements the {\em \SCD-broadcast} communication abstraction
in $\CAMP_{n,t}[t<n/2]$. Moreover, it requires $O(n^2)$
messages per invocation of $\scbroadcast()$. 
\end{theorem}

\begin{proofT}
The proof follows from
Lemma~\ref{lemma-broadcast-validity} (Validity), 
Lemma~\ref{lemma-broadcast-integrity} (Integrity), 
Lemma~\ref{lemma-broadcast-sc-ordering} (MS-Ordering), 
Lemma~\ref{lemma-broadcast-termination1} (Termination-1), and
Lemma~\ref{lemma-broadcast-termination2} (Termination-2).

The  $O(n^2)$ message complexity comes from the fact that,
due to the predicates of line~\ref{SCD-from-MP-05} and~\ref{SCD-from-MP-06},
each application message $m$ is forwarded at most once by each process
(line~\ref{SCD-from-MP-11}).
\renewcommand{\toto}{theorem:SCD}    
\end{proofT}

The next corollary follows from (i)
Theorems~\ref{theorem:proof-snapshot}  and~\ref{theorem:SCD}, and (ii) 
the fact that the constraint
$(t<n/2)$ is an upper bound on the number of faulty processes
to build a read/write register (or snapshot object)~\cite{ABD95}.

\begin{corollary}
\label{coro:sc-optimal}
Algorithm~{\em\ref{algo:Sc-broadcast-in-RW}} is resiliency optimal. 
\end{corollary} 

\section{Building an MWMR atomic register on top
  of $\CAMP_{n,t}[\mbox{\SCD-broadcast}]$}
\label{sec:from-SCD-to-SWMR-MWMR}

This appendix shows the genericity dimension of
Algorithm~\ref{algo:snapshot-from-SCD}.
It presents trivial simplifications of it, which build MWMR
atomic registers and MWMR  sequentially consistent registers.

\subsection{The algorithm}
\label{sec:SWMR-RW}
Let $\REG$ denote the MWMR atomic read/write register that is built.
The algorithm that builds it is a trivial simplification of the
snapshot Algorithm~\ref{algo:snapshot-from-SCD},
namely its projection on a single MWMR atomic register.

$\REG$  is now locally represented 
by a  local variable $reg_i$ and the associated timestamp $ts_i$
initialized to $\langle 0,-\rangle$.  The message sent at
Line~\ref{Snap-from-SC-09} is now \WRITE $(v,\langle ts_i.date_i+1,i\rangle)$,
and the predicate of line~\ref{Snap-from-SC-11} simplifies to  
``there are messages \WRITE())''.

\subsection{Proof of the algorithm}

The proof is a simplified version of the proof of
Theorem~\ref{theorem:proof-snapshot}. For self-completeness,
we give here its full proof even if some parts of it are
``cut-and-paste'' of parts of proofs given in Section~\ref{sec:proof-snapshot}. 
As in that section, 
let us associate a timestamp $ts(\op)$ with each operation $\op()$ 
as follows (this is the place where the proof is simplified 
with respect to a snapshot object). 
\begin{itemize}
\item Case  $\op()=\wwrite(v)$. Let $p_i$ be the invoking process;
  $ts(\op)$ is the timestamp of $v$ 
  as defined by $p_i$ at line~\ref{Snap-from-SC-09}, i.e.,
  $\langle ts_i.date+1, i\rangle$. 

\item Case $\op()=\rread()$. Let $w$ be the value returned by
  the read;  $ts(\op)$ is then the timestamp associated with $w$ at
  line~\ref{Snap-from-SC-15} by its writer. 
\end{itemize}

Let $\op1$ and $\op2$ be any two operations. The relation
$\prec$ on the whole set of operations is defined as follows:
 $\op1 \prec \op2$ if $\op1$ terminated before $\op2$ started. 
It is easy to see that $\prec$ is a real-time-compliant partial order 
on all the operations.

The reader can easily check that the statement and the  proof of
Lemma~\ref{lemma:snapshot-liveness} (applied to the termination of read and
write operations), and Lemma~\ref{lemma:snapshot-write-ordering} (applied
to the total order on the write operations, compliant with both
the sequential specification of a register, and their real-time occurrence
order)  remain valid for the algorithm suited to an MWMR atomic read/write
register.  The next lemma addresses the read operations (which are simpler
to manage than snapshot operations).

\begin{lemma}
 \label{lemma:one-reg-memory-safety}
 The read/write register $\REG$ is linearizable.
\end{lemma}

\begin{proofL}
Let us now insert each read operation in the previous (real time compliant)
total order  as follows.

Let $\rread1()$ be a read operation  whose timestamp is
$\langle date1, i \rangle$. This operation is inserted just after the
write operation  $\wwrite1()$  that has the same timestamp
(this write wrote the value  read by $\rread1()$).
Let us remark that, as  $\rread1()$ obtained the value timestamped
$\langle date1, i \rangle$,  it  did not terminate  before  $\wwrite1()$
started.  It follows that the insertion of $\rread1()$ into the total order
cannot violate the real-time order between $\wwrite1()$ and  $\rread1()$.

Let us consider the operation  $\wwrite2()$ that follows
$\wwrite1()$ in the write total order. 
If $\rread1() \prec \wwrite2()$, the placement of $\rread1()$
in the total order is real-time-compliant.
If  $\neg(\rread1() \prec \wwrite2())$, due to the timestamp obtained by 
$\rread1()$, we cannot  have $ \wwrite2()\prec \rread1()$.
It follows that in this  case also, the placement of $\rread1()$
in the total order is real-time-compliant.

Finally, let us consider two read operations $\rread1()$ and $\rread2()$
which have the same timestamp  $\langle date, i \rangle$ (hence, they read
from the same write operation, say $\wwrite1()$). 
Both are inserted after  $\wwrite1()$ in the order of their invocations (if 
$\rread1()$ and $\rread2()$ started simultaneously, they are  inserted
according to the order on the identities of the processes that invoked them).
Hence,  the read and write operations are linearizable, which concludes
the proof of the lemma.
\renewcommand{\toto}{lemma:one-reg-memory-safety}
\end{proofL}

\begin{theorem}
\label{theorem:proof-one-register}
The read/write register $\REG$
is an {\em MWMR} atomic read/write register.
\end{theorem}

\begin{proofT}
The proof follows from Lemma~\ref{lemma:snapshot-liveness},
Lemma~\ref{lemma:snapshot-write-ordering},
and Lemma~\ref{lemma:one-reg-memory-safety}. 
\renewcommand{\toto}{theorem:proof-one-register}
\end{proofT}

\subsection{The case of an SWMR atomic register}
When the register $\REG$ can be written by a single process (say $p_k$), 
the algorithm simplifies. The timestamps disappear at all processes, and
as only the writer $p_k$ can invoke $\REG.\wwrite()$, 
it  manages a simple  date $date_k$ (which is actually
a sequence number). The modifications are:
\begin{itemize}
\item  Line~\ref{Snap-from-SC-09}   becomes:
$date_k\leftarrow date_k+1$; $\scbroadcast$ \WRITE $(v,date_k)$. 

\item 
The lines~\ref{Snap-from-SC-11}-\ref{Snap-from-SC-17} become:
{\small
\begin{tabbing}
aaaaaaaaaaaaaaaaa\=aa\=aaa\=aaaaaa\=\kill
  
\> {\bf if} \= (there are messages \WRITE()) \\

\>\> {\bf then} \= {\bf let} $date$ be the maximal date in
                   the messages \WRITE() received;\\

\>\>\>
   $reg_i\leftarrow$ the value associated with $date$\\
 \> {\bf end if}.
 
\end{tabbing}
}
\end{itemize}
Let us remark that, due to the Boolean $\done_k$, the writer $p_k$
scd-delivers message sets containing at most one message \WRITE().

\subsection{On sequentially consistency}

\paragraph{The case of an MWMR sequentially consistent register}
As indicated in the Introduction, sequential consistency was introduced
in~\cite{L79}. It is atomicity minus the  requirement stating that
``if  an operation $op1$ terminates before an operation $op2$ starts, then 
$op1$ must appear before $op2$ in the sequence of the read and write
operations''. As noticed in~\cite{R02}, sequential consistency can be
seen as a weakened form of atomicity, namely lazy linearizability.
The composition of sequentially consistent registers is investigated
in~\cite{PMMJ16}. The algorithm for sequential consistency
presented in~\cite{PMMJ16} and Algorithm \ref{algo:Sc-broadcast-in-RW}
are based on similar principles. 
The constraint $(t<n/2)$ is also  a necessary and sufficient condition to
implement  a sequentially consistent read/write register
in ${\CAMP}_{n,t}[\emptyset]$. 

The reader can check that an algorithm building a  a sequentially
consistent MWMR read/write register can easily be obtained from
Algorithm~\ref{algo:snapshot-from-SCD} as simplified in
Section~\ref{sec:SWMR-RW}.
One only needs to suppress the synchronization messages \SYNC()
which ensure the compliance with respect to real-time.
The concerned lines are  lines~\ref{Snap-from-SC-01}-\ref{Snap-from-SC-03}
(read synchronization), and lines~\ref{Snap-from-SC-05}-\ref{Snap-from-SC-07}
(write synchronization). In a simple way, this  shows the versatility
dimension of Algorithm~\ref{algo:snapshot-from-SCD}.

\paragraph{From sequential consistency to atomicity}
Given a sequentially consistent snapshot object,
Algorithm~\ref{algo:Sc-broadcast-from-snapshot} builds the
\SCD-broadcast communication abstration.  (As the reader can check,
this follows from the fact that, when looking at its proof, this
algorithm relies only on the fact that the operations on the snapshot
object can be totally ordered.)  Hence, using on top of it the
\SCD-broadcast-based Algorithm~\ref{algo:snapshot-from-SCD}, we obtain
an atomic snapshot object.  It follows that, thanks to SCD-broadcast,
the algorithms presented in the paper allow a sequentially consistent
snapshot object to be transformed into an atomic snapshot object
(and it is known that --differently from sequential consistent objects-- 
atomic objects are composable for free~\cite{HW90}). 


\begin{thebibliography}{}

\end{thebibliography}


\begin{thebibliography}{99}


{\small

\bibitem{AADGMS93} 
Afek Y., Attiya H., Dolev D., Gafni E., Merritt M. and Shavit N., 
Atomic snapshots of shared memory. 
{\em Journal of the ACM}, 40(4):873-890 (1993)
    

\bibitem{ANBHK95}
Ahamad M.,  Neiger G.,  Burns J.E.,  Hutto P.W., and Kohli P. 
Causal memory: definitions, implementation and programming. 
{\it Distributed Computing}, 9:37-49 (1995)   

\bibitem{A94}
Anderson J., 
Multi-writer composite registers.
{\it Distributed Computing}, 7(4):175-195 (1994)

\bibitem{A00}
Attiya H.,
Efficient and robust sharing of memory in message-passing systems.
{\it Journal of Algorithms}, 34:109-127 (2000) 


\bibitem{ABD95}
Attiya H., Bar-Noy A. and Dolev D.,
Sharing memory robustly in  message passing systems. 
{\it Journal of the ACM}, 42(1):121-132 (1995)


\bibitem{AR98}
Attiya H. and Rachmann O.,
Atomic snapshots in $O(n\log n)$ operations. 
{\it SIAM Journal of Computing},  27(2):319-340 (1998)




\bibitem{AW94}
Attiya H. and  Welch J.L., 
Sequential consistency versus linearizability.
{\it ACM Transactions on Computer Systems}, 12(2):91-12 (1994)

\bibitem{AW04}
Attiya H. and Welch J.L.,  
{\it Distributed computing: fundamentals, simulations  and advanced  topics}, 
(2d  Edition),  Wiley-Interscience, 414  pages (2004) 

\bibitem{BJ87}
Birman K.  and  Joseph T.
Reliable communication in the presence of failures.
{\it ACM Transactions on Computer Systems}, 5(1):47--76 (1987)


\bibitem{CT96}
Chandra T. and Toueg S., 
Unreliable failure detectors for reliable distributed systems. 
{\em Journal of the ACM}, 43(2):225-267 (1996)


\bibitem{DFRR16}
Delporte-Gallet  C., Fauconnier H., Rajsbaum S., and Raynal M., 
Implementing snapshot objects on top of 
crash-prone asynchronous message-passing systems.
{\it Proc.  16th Int'l Conference on Algorithms and Architectures for
Parallel Processing (ICA3PP'16)}, Springer LNCS 10048,  pp.~341--355 (2016)


\bibitem{DGLV10} 
Dutta P.,  Guerraoui R.,   Levy R., and  Vukolic M., 
Fast  access to distributed atomic memory.
{\it SIAM Journal of  Computing}, 39(8):3752-3783 (2010)


\bibitem{E05}
Ellen F.,
How hard is it to take a snapshot? 
{\it  Proc. 31th Conference on Current Trends in 
Theory   and  Practice  of   Computer  Science (SOFSEM'05)},
 Springer 3381, pp.~27-35 (2005)


\bibitem{EFR07}
Ellen F., Fatourou P., and  Ruppert E., 
Time lower bounds for implementations of multi-writer snapshots.
{\it Journal of the ACM}, 54(6), 30 pages (2007)


 


\bibitem{FLP85}
Fischer M.J., Lynch N.A. and Paterson M.S.,
Impossibility of distributed consensus with one faulty process. 
{\em Journal of the ACM}, 32(2):374-382 (1985)

\bibitem{FM03}
Fischer M.J. and Merritt M.,
Appraising two decades of distributed computing theory research.
{\it Distributed Computing},  16(2-3):239-247 (2003)
  

\bibitem{HNS16}
Hadjistasi Th.,  Nicolaou N., and  Schwarzmann A.A., 
Oh-RAM! One and a half round read/write atomic memory.
{\it Brief announcement.  Proc. 35th ACM Symposium on  Principles of 
Distributed Computing (PODC'16)}, ACM Press, pp.~353-355 (2016)



\bibitem{HW90} 
Herlihy  M. P.  and Wing J. M.,  Linearizability: a correctness
condition  for concurrent  objects.  
{\it  ACM Transactions  on Programming Languages and Systems}, 
12(3):463-492 (1990)



\bibitem{IR12}
Imbs D. and Raynal M., 
Help when  needed, but no more:  efficient read/write partial snapshot. 
{\it  Journal of Parallel and Distributed Computing}, 72(1):1-12 (2012)

\bibitem{ICMT94}
Inoue I.,  Chen W., Masuzawa T. and Tokura N.,
Linear time snapshots using multi-writer multi-reader registers.
{\it Proc.  8th Int'l Workshop on Distributed Algorithms (WDAG'94)}, 
Springer LNCS 857, pp.~130-140 (1994)


\bibitem{J05}
Jayanti P., 
An optimal multiwriter snapshot algorithm. 
{\it Proc. 37th ACM Symposium on Theory of Computing (STOC'05)}, 
ACM Press, pp. 723-732 (2005)


\bibitem{JFC08}
Jim\'enez E.,  Fern\'andez A.,  and Cholvi V.,  
A parameterized algorithm that implements sequential, causal, and cache memory 
consistencies. 
{\it Journal of Systems and Software},  81(1):120-131 (2008)


\bibitem{K56}
Kramer S. N., 
{\it History Begins at Sumer: Thirty-Nine Firsts in Man's Recorded History}.
University of Pennsylvania Press, 416 pages, ISBN 978-0-8122-1276-1 (1956)

\bibitem{L79}
Lamport L., 
How to make a multiprocessor computer that correctly executes
multiprocess programs.
{\it IEEE Transactions on Computers}, C28(9):690--691 (1979)

  
\bibitem{L86}
Lamport L.,
On interprocess communication, Part I: basic formalism. 
{\it Distributed Computing}, 1(2):77-85 (1986) 



\bibitem{L96}
Lynch N. A.,
{\it Distributed algorithms}.
Morgan Kaufmann Pub., San Francisco (CA), 872 pages, ISBN 1-55860-384-4 (1996)

\bibitem{M86} 
Misra  J., Axioms for memory access  in asynchronous hardware systems.   
{\it ACM  Transactions  on Programming  Languages and  Systems},
8(1):142-153 (1986)

\bibitem{MPRJ17} 
Most\'efaoui A., P\'etrolia M.,  Raynal M., and Jard Cl., 
Atomic read/write memory in signature-free  
 Byzantine asynchronous message-passing systems.
{\it Springer  Theory of  Computing Systems} (2017)
DOI: 10.1007/s00224-016-9699-8




\bibitem{MR16-podc}
Most\'efaoui A. and Raynal M.,
Two-bit messages are sufficient to implement  atomic read/write 
registers in crash-prone  systems. 
{\it Proc. 35th ACM Symposium on  Principles of 
Distributed Computing (PODC'16)}, ACM Press, pp.~381-390 (2016)

\bibitem{PMMJ16}
Perrin M., Most\'efaoui A., P\'etrolia M., and Jard Cl., 
On composition and implementation of sequential consistency.
{\it Proc. 30th Int'l  Symposium on  Distributed Computing (DISC'16)},
Springer LNCS 9888, pp.~284-297 (2017)


\bibitem{R02}
Raynal M., 
Sequential consistency as lazy linearizability.  
{\it Brief announcement.
  Proc. 14th  ACM Symposium on Parallel Algorithms and Architectures
(SPAA'02)}, ACM press, pp.~151-152, (2002)

\bibitem{R10}
Raynal M.,
{\it Communication and agreement abstractions for fault-tolerant 
asynchronous distributed systems.} 
Morgan \& Claypool Publishers, 251 pages, ISBN 978-1-60845-293-4 (2010)


\bibitem{R13}
Raynal M.,
{\it Distributed algorithms for message-passing systems}. 
Springer, 510 pages, ISBN 978-3-642-38122-5  (2013)


\bibitem{R13-1}
Raynal M.,
{\it Concurrent programming: algorithms, principles and  foundations}.
Springer,  515 pages, ISBN 978-3-642-32026-2 (2013) 


\bibitem{RST91}
Raynal M., Schiper A., and Toueg S., 
The causal ordering abstraction and a simple way to implement it.
{\it Information Processing Letters}, 39:343-351 (1991)
  

\bibitem{R08}
Ruppert E., 
Implementing shared registers in asynchronous message-passing systems. 
{\it Springer Encyclopedia of Algorithms}, pp.~400-403 (2008)


\bibitem{T36}
Turing A.M.,
On computable numbers with an application to the Entscheidungsproblem.
{\it Proc. of the London Mathematical Society}, 42:230-265 (1936)


\bibitem{V12}
Vukolic M., 
{\it Quorum systems, with applications to storage and consensus}. 
Morgan \& Claypool Publishers, 132 pages, ISBN 978-1-60845-683-3 (2012)

}

\end{thebibliography}
\end{document}